\documentclass[aps,amsmath,amssymb,twocolumn,showpacs,byrevtex,superscriptaddress]{revtex4}
\usepackage{graphicx}

\begin{document}

\title{From caging to Rouse dynamics in polymer melts with intramolecular barriers:\\
a critical test of the Mode Coupling Theory}

%
%

\author{Marco Bernabei}
\affiliation{Donostia International Physics Center, Paseo Manuel de Lardizabal 4,
E-20018 San Sebasti\'{a}n, Spain.}
\author{Angel J. Moreno}
\email[Corresponding author: ]{wabmosea@ehu.es}
\affiliation{Centro de F\'{\i}sica de Materiales (CSIC, UPV/EHU) and Materials Physics Center MPC, 
Paseo Manuel de Lardizabal 5, E-20018 San Sebasti\'{a}n, Spain.}
\author{Emanuela Zaccarelli}
\affiliation{Dipartimento di Fisica and CNR-ISC, 
Universit\`{a} di Roma La Sapienza,\\ Piazzale Aldo Moro 2, I-00185, Roma, Italy} 
\author{Francesco Sciortino}
\affiliation{Dipartimento di Fisica and CNR-ISC, 
Universit\`{a} di Roma La Sapienza,\\ Piazzale Aldo Moro 2, I-00185, Roma, Italy} 
\author{Juan Colmenero}
\affiliation{Donostia International Physics Center, Paseo Manuel de Lardizabal 4,
E-20018 San Sebasti\'{a}n, Spain.}
\affiliation{Centro de F\'{\i}sica de Materiales (CSIC, UPV/EHU) and Materials Physics Center MPC, 
Paseo Manuel de Lardizabal 5, E-20018 San Sebasti\'{a}n, Spain.}
\affiliation{Departamento de F\'{\i}sica de Materiales, Universidad del Pa\'{\i}s Vasco (UPV/EHU),\\
Apartado 1072, E-20080 San Sebasti\'{a}n, Spain.}

\begin{abstract}

By means of computer simulations and solution of the equations of the Mode Coupling Theory (MCT), 
we investigate the role of the intramolecular barriers on several dynamic aspects of non-entangled polymers. 
The investigated  dynamic range extends from the caging regime characteristic of glass-formers
to the relaxation of the chain Rouse modes. We review our recent work on this question,
provide new results and critically discuss the limitations of the theory.
Solutions of the MCT for the structural relaxation reproduce qualitative trends of simulations for weak
and moderate barriers. However a progressive discrepancy is revealed as
the limit of stiff chains is approached. This disagreement does not seem related with 
dynamic heterogeneities, which indeed are not enhanced by increasing barrier strength.
It is not connected either with the breakdown of the convolution approximation
for three-point static correlations, which retains its validity for stiff chains.
These findings suggest the need of an improvement of the MCT equations for polymer melts.
Concerning the relaxation of the chain degrees of freedom, MCT provides a microscopic basis 
for time scales from chain reorientation down to the caging regime.
It rationalizes, from first principles, the observed devations from the Rouse model
on increasing the barrier strength. These include anomalous 
scaling of relaxation times, long-time plateaux, and non-monotonous wavelength dependence 
of the mode correlators.

\end{abstract}
\date{\today}
\pacs{64.70.pj, 64.70.qj, 61.20.Ja, 83.10.Mj, 83.10.Rs, 83.80.Sg}
\maketitle

\begin{center}
{\bf I. INTRODUCTION}
\end{center}

The different dynamic processes present in amorphous polymers cover a extremely broad range
of characteristic time scales, spanning from about 100 femtoseconds  up to years. 
There are two main reasons for this. First, polymers are usually good glass-formers,
which inherently exhibit a dramatic increase of the viscosity and structural ($\alpha$-) relaxation times
on approaching the glass transition temperature $T_{\rm g}$. As in non-polymeric glass-formers,
localized dynamic processes are also present below $T_{\rm g}$ \cite{McCrum:Book1967}.
Second, their macromolecular character
introduces relaxation processes related to the dynamics of the internal chain degrees of freedom.
In the case of low-molecular weight, nonentangled, polymer chains a sublinear increase (Rouse-like) arises
in the mean squared displacement prior to the linear diffusive regime.
In the case of high-molecular weight, strongly entangled, chains 
further sublinear regimes are found between the Rouse and linear regimes,
which are usually interpreted in terms of reptation dynamics
\cite{DoiEdwards:Book1986,DeGennes:Book1979,McLeish:AdvPhys2002}.
Such processes are inherent to chain connectivity, and extend over more time decades 
on increasing chain length. This broad time window for chain dynamics is observed
even for temperatures far above $T_{\rm g}$, 
when the structural relaxation extends over just a few picoseconds.

Another particular ingredient of polymer systems is that, apart from fast librations 
or methyl group rotations \cite{Colmenero:ProgPolSci2005}, 
every motion involves jumps over carbon-carbon rotational barriers and/or chain conformational changes. 
The corresponding map of relaxation processes is largely influenced by the barrier strength.
Intramolecular barriers play a decisive role
in, e.g., crystallization \cite{Meyer:Macrom2002,Vettorel:PRE2007}, 
adsorption onto surfaces \cite{Semenov:EPJE2002,Ubbink:JCP2004},
viscoelastic properties \cite{Morse:Macrom1998}, or phase behavior of block copolymers \cite{Singh:Macrom1994}.
Models for semiflexible and stiff polymers are of great interest in biophysics, 
since they can be applied to many important biopolymers 
as DNA, rodlike viruses, or actin filaments
\cite{Bustamante:Science1994,Kas:BiophysJ1996,Ober:Science2000}.
Thus, an understandig of the role of the intramolecular barriers on structural and dynamic
properties of polymer systems is of practical as well as of fundamental interest
in many fields of research.

A possible theroretical approach to this problem is provided by the Mode Coupling Theory (MCT) 
\cite{Gotze:Book2009}. 
MCT introduces a closed set of coupled Mori-Zwanzig equations for the time
dependence of density correlators. 
Static correlations enter the memory kernel as external input. Since the former can be related to
the interaction potential through liquid state theories, MCT constitutes a first-principle theory 
for slow dynamics in complex systems.
MCT has been developed over the last years to include systems
with intramolecular structure (see e.g., \cite{Chong:PRE1998,Chong:PRE2002,Chong:PRE2004}). 
This includes the approach of Chong and co-workers for
simple polymer melts \cite{Chong:PRL2002,Chong:PRE2007},
based on the polymer reference interaction site model 
(PRISM) \cite{Schweizer:AdvChemPhys1997}
for the static correlations. This approach was applied to 
the specific case of fully-flexible chains \cite{Chong:PRL2002,Chong:PRE2007},
i.e., without intramolecular barriers. 
A major success was the derivation, from first-principles, 
of the scaling laws predicted by the phenomenological 
Rouse model \cite{DoiEdwards:Book1986} for chain dynamics in nonentangled polymer melts.
Likewise, it provided a unified microscopic description of both chain dynamics and
the structural relaxation associated to the glass transition \cite{Chong:PRL2002,Chong:PRE2007}.

Some of us have recently performed a systematic computational investigation
of the role of intramolecular barriers on the glass transition
in polymer systems \cite{Bernabei:PRL2008,Bernabei:JCP2009}. 
Starting from fully-flexible bead-spring chains, we introduced  stiffness by implementing 
intramolecular barriers with tunable bending and torsion terms. 
In Ref.~\cite{Bernabei:JCP2009} we discussed the glass transition within the framework of the MCT
for polymer melts, comparing simulations with numerical solutions of the MCT equations,
in the long-time limit, for a broad range of barrier strength. This was possible since 
the quality of the PRISM approximations observed for fully-flexible chains \cite{Aichele:PRE2004}
was not affected at all by the introduction of internal barriers 
in all the investigated range \cite{Bernabei:JCP2009}. 
Numerical solutions  reproduced  
trends in the nonergodicity parameters and MCT critical temperatures
for weak and moderate barriers. However, strong discrepancies were observed 
on approaching the limit of stiff chains \cite{Bernabei:JCP2009}.

In this article we briefly summarize the main points of Refs.~\cite{Bernabei:PRL2008,Bernabei:JCP2009}
and present extensive new results. Thus, we solve the time-dependent MCT equations 
for density correlators, and compare simulation and theoretical trends in $\alpha$-relaxation times.
We critically discuss the limitations of the theory by analyzing the accuracy of the assumed
approximations. We find that dynamic heterogenities, static three-point correlations,
and chain packing effects not accounted by MCT, do not play a major role on increasing the barrier
strength. Indeed their effects seem to be weaker that in the case of fully-flexible chains. 
The reason for the observed discrepancies between simulation and theory for very stiff chains
remains to be understood. 

We also present here a systematic investigation on the effect of intramolecular barriers
on the internal chain dynamics of nonentangled polymers.
We analyze correlators for the chain normal modes (Rouse modes)
and for bond reorientation. The simulations reveal strong deviations from the Rouse model
on increasing chain stiffness. These include anomalous scaling of relaxation times \cite{Steinhauser:JCP2009},
long-time plateaux, and nonmonotonous wavelength dependence of the mode correlators.
We show that these anomalous dynamic features 
are reproduced by the corresponding MCT equations for the Rouse modes.
This generalizes the analysis
of Ref.~\cite{Chong:PRE2007}, which was limited to fully-flexible chains, to polymers
with intramolecular barriers of arbitrary strength. 
Thus, beyond usual phenomenological models for chain dynamics,
MCT provides a unified microscopic picture  
down to time scales around and before the $\alpha$-process \cite{noteTg}.

The article is organized as follows. In Section II we describe the model and give simulation details.
In Section III we compare simulation results with MCT solutions
for several dynamic correlators probing structural relaxation and chain dynamics.
In Section IV we discuss the possible origin of the observed deviations from MCT predictions.
Conclusions are given in Section V.

\begin{center}
{\bf II. MODEL AND SIMULATION DETAILS}
\end{center}

We have performed molecular dynamics (MD) simulations of a bead-spring  model 
with tunable  intramolecular barriers. All chains consist of $N_{\rm m} = 10$
identical monomers of mass $m =1$.
Non-bonded interactions between monomers are given  by a corrected soft-sphere potential
\begin{equation}
V(r) = 4\epsilon[(\sigma/r)^{12} - C_0 + C_2(r/\sigma)^{2}],
\label{eq:potsoft}
\end{equation}
where $\epsilon=1$ and $\sigma=1$. The potential $V(r)$ is set to zero for $r \geq c\sigma$, with $c = 1.15$.
The values $C_0 = 7c^{-12}$ and $C_2 = 6c^{-14}$  guarantee continuity of  potential and forces at 
the cutoff distance $r=c\sigma$. 
The potential $V(r)$ is purely repulsive. It does not show local minima within the interaction range $r<c\sigma$. 
Thus, it drives dynamic arrest only through packing effects. 
Chain connectivity is introduced by means of  a finitely-extensible nonlinear elastic (FENE)  
potential \cite{Grest:PRA1986,Bennemann:EPJB1999} between consecutive monomers:
\begin{equation}
V_{\rm FENE}(r) = -\epsilon K_{\rm F}R_0^2 \ln[ 1-(R_0\sigma)^{-2}r^2 ],
\label{eq:potfene}
\end{equation}
where $K_{\rm F}=15$ and $R_0=1.5$. The superposition of potentials (\ref{eq:potsoft}) and (\ref{eq:potfene}) 
yields an effective bond potential for consecutive monomers
with a sharp minimum at $r\approx 0.985$, which makes bond crossing impossible.

Intramolecular barriers are implemented by means of the combined bending and torsional potentials
proposed by Bulacu and van der Giessen in Refs.~\cite{Bulacu:JCP2005,Bulacu:PRE2007}.
The bending potential $V_{\rm B}$ acts on three consecutive monomers along the chain and is defined as 
\begin{equation}
V_{\rm B}(\theta_i) = (\epsilon K_{\rm B}/2 )(\cos\theta_i - \cos\theta_0)^2,
\label{eq:potben}
\end{equation}
where $\theta_i$ is the bending angle between consecutive monomers $i-1$, $i$ and $i+1$ (with $2 \leq i \leq N_{\rm m}-1$).
We use $\theta_0 = 109.5^{\rm o}$  for the equilibrium bending angle.
The torsional  potential $V_{\rm T}$ constrains the dihedral angle $\phi_{i,i+1}$.
The latter is defined for the consecutive monomers  
$i-1$, $i$, $i+1$ and $i+2$ (with $2 \leq i \leq N_{\rm m}-2$), as the angle between the two planes defined by the sets
($i-1$, $i$, $i+1$) and ($i$, $i+1$, $i+2$). The form of the torsional potential is
\begin{eqnarray}
V_{\rm T}(\theta_{i},\theta_{i+1},\phi_{i,i+1}) = \hspace{2 cm} \nonumber \\
\epsilon K_{\rm T}\sin^3 \theta_{i} \sin^3 \theta_{i+1} \sum_{n=0}^3 a_n \cos^n \phi_{i,i+1}.
\label{eq:pottor}
\end{eqnarray}
The values of the coefficients $a_n$ are $a_0=3.00$, $a_1=-5.90$, $a_2=2.06$, and $a_3=10.95$ \cite{Bulacu:JCP2005,Bulacu:PRE2007}.
The torsional potential depends both on the dihedral angle $\phi_{i,i+1}$  
and on the bending angles $\theta_i$ and $\theta_{i+1}$. As noted in Refs.~\cite{Bulacu:JCP2005,Bulacu:PRE2007}, the functional
form (\ref{eq:pottor}) avoids numerical instabilities arising when two consecutive bonds align,
without the need of imposing rigid constraints on the bending angles.

In the following, temperature $T$, time $t$, distance, wave vector $q$, and monomer density $\rho$ 
are given respectively in units of $\epsilon/k_B$ (with $k_B$ the Boltzmann constant), 
$\sigma(m/\epsilon)^{1/2}$, $\sigma$, $\sigma^{-1}$, and $\sigma ^{-3}$.
We investigate, at fixed monomer density $\rho = 1.0$, the  temperature dependence of the dynamics 
for different values of the bending and torsion strength,
$(K_{\rm B}$,$K_{\rm T}) =$ (0,0), (4,0.1), (8,0.2), (15,0.5), (25,1), (25,4), and (35,4), 
covering a broad dynamic range from the
caging characteristic time to the relaxation time of the slowest Rouse mode.
We investigate typically 8-10 different temperatures for each set of values  $(K_{\rm B}$,$K_{\rm T})$.
Additional numerical details can be found in Refs.~\cite{Bernabei:PRL2008,Bernabei:JCP2009}.

The investigated range of barrier strength corresponds to a strong variation of the chain stiffness.
This can be quantified by the average end-to-end radii, $R_{\rm ee}$, of the chains. 
Thus, for the representative values 
$(K_{\rm B},K_{\rm T})= (0,0)$, (8,0.2), (25,1) and (35,4), which cover the range from fully-flexible chains
to the stiffest investigated chains, we find $R_{\rm ee} = 3.6$, 4.7, 5.5, and 6.5 
at the respective lowest investigated temperature.

\begin{center}
{\bf III. RESULTS: SIMULATIONS vs. THEORY}
\end{center}

\begin{figure}
\begin{center}
\includegraphics[width=0.85\linewidth]{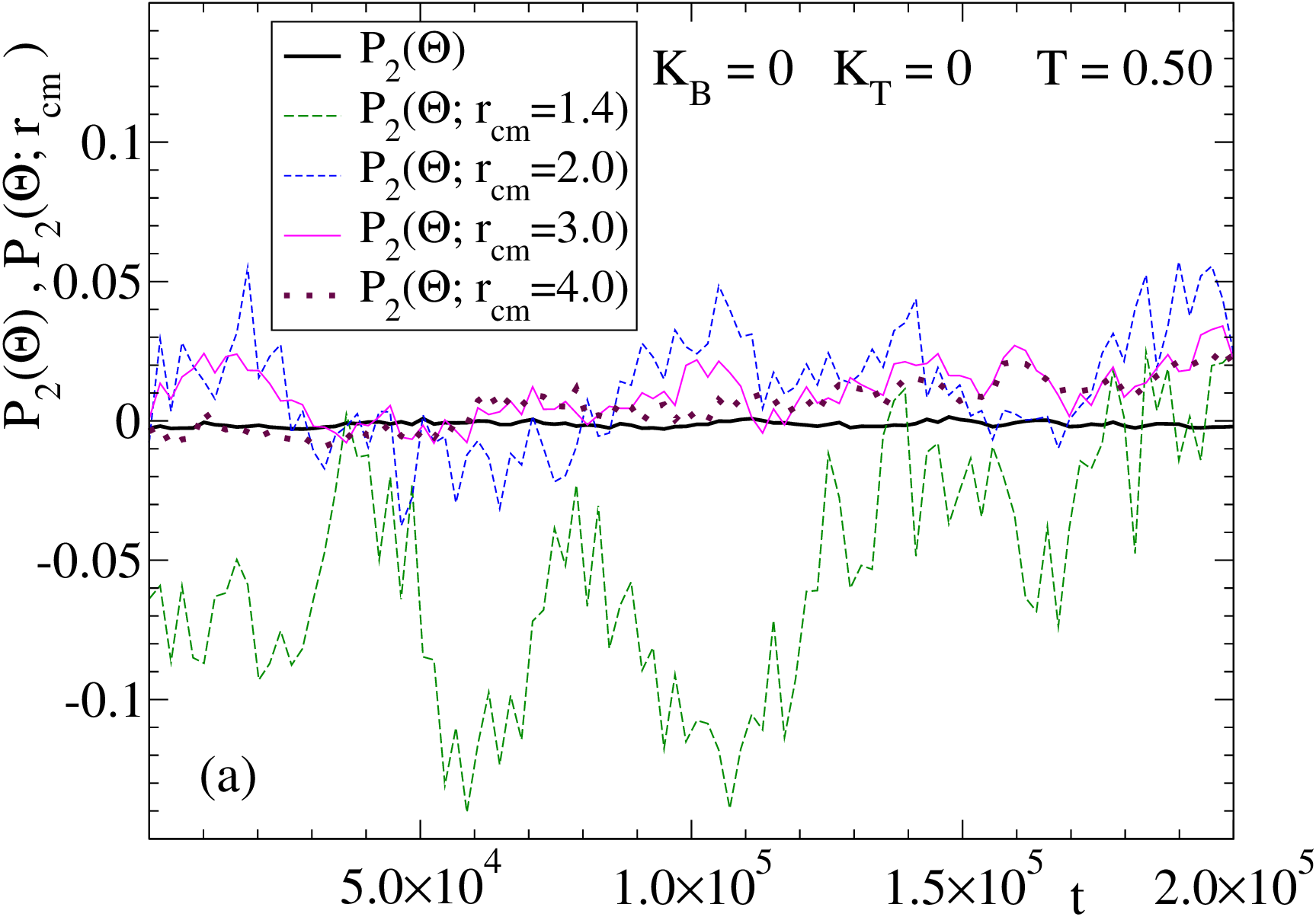}
\newline
\newline
\includegraphics[width=0.85\linewidth]{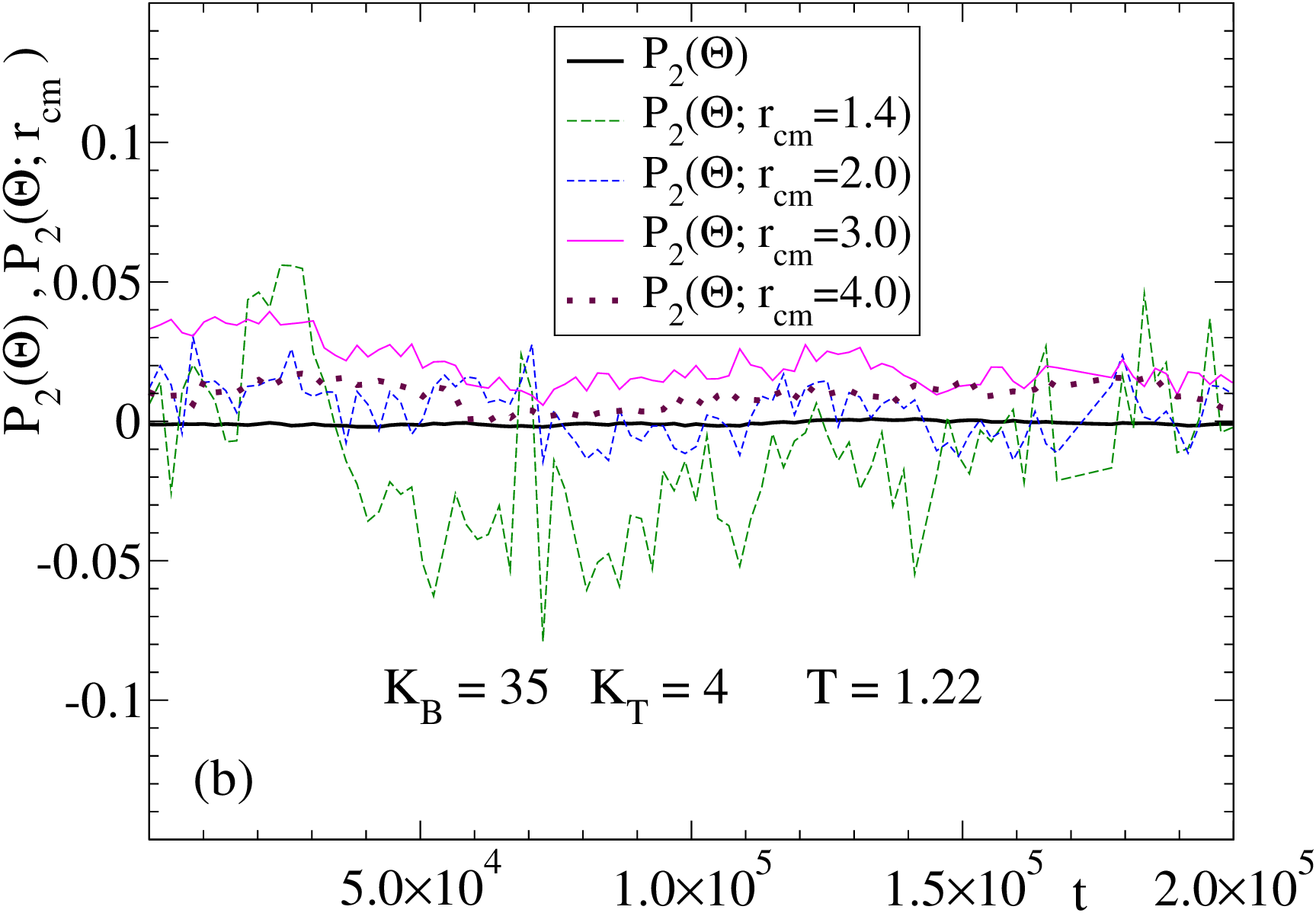}
\newline
\caption{Time evolution of the global and local
orientational parameter (see text) 
at two selected temperatures, for fully-flexible chains (a) and stiff chains (b) with 
$(K_{\rm B}$,$K_{\rm T}) =(35,4)$.}
\label{fig:nematic}
\end{center}
\end{figure}

Before addressing the dynamic aspects of the present system, we want to stress
that the investigated state points correspond to isotropic phases.
We do not observe signatures of {\it global} orientational order induced by chain stiffness
for the investigated state points. 
Thus, by measuring the quantity  $P_2 (\Theta) = (3\langle\cos ^2\Theta\rangle -1)/2$, 
where $\Theta$ is the angle between the end-to-end vectors of two chains, and averaging it 
over all pairs of distinct chains,  we obtain in all cases values $|P_2 (\Theta)| < 3\times 10^{-3}$.
This is illustrated in Fig.~\ref{fig:nematic}, which shows the time evolution of $P_2 (\Theta)$ 
along a typical simulation window,
both for fully-flexible chains, $(K_{\rm B}$,$K_{\rm T}) =$ (0,0), 
and for the stiffest investigated chains, $(K_{\rm B}$,$K_{\rm T}) =$ (35,4).

{\it Local} orientational order is also negligible. This is evidenced by computing
a similar correlator $P_2 (\Theta ;r_{\rm cm})$. In this case the average is performed only 
over pairs of distinct chains for which the distance between their respective 
centers-of-mass is less than $r_{\rm cm}$.
Fig.~\ref{fig:nematic} displays, for the former cases of fully-flexible and stiff chains, 
data of $P_2 (\Theta ;r_{\rm cm})$ for several values of $r_{\rm cm}$.
Negligible values of $P_2 (\Theta ;r_{\rm cm})$ are obtained for $r_{\rm cm} \geq 2.0$. 
Thus, the time average
over the simulation time window, $t_{\rm sim}$, provides values 
$|\langle P_2 (\Theta ;r_{\rm cm} \geq 2.0)\rangle_{\rm t_{\rm sim}}| < 0.02$.
By comparing both panels we conclude that chain stiffness does not induce a significant 
increase, if any, of local orientational order in the investigated systems.
Weak local orientational order 
$|\langle P_2 (\Theta ;r_{\rm cm})\rangle_{\rm t_{\rm sim}}| \lesssim 0.1$ is observed
only for very small interchain distances (see data for $P_2 (\Theta ;r_{\rm cm} = 1.4)$).
Again, the introduction
of chain stiffness does not induce clear changes in the orientational order at this length scale.

\begin{center}
{\bf A. Structural relaxation}
\end{center}

\begin{figure}
\begin{center}
\includegraphics[width=0.86\linewidth]{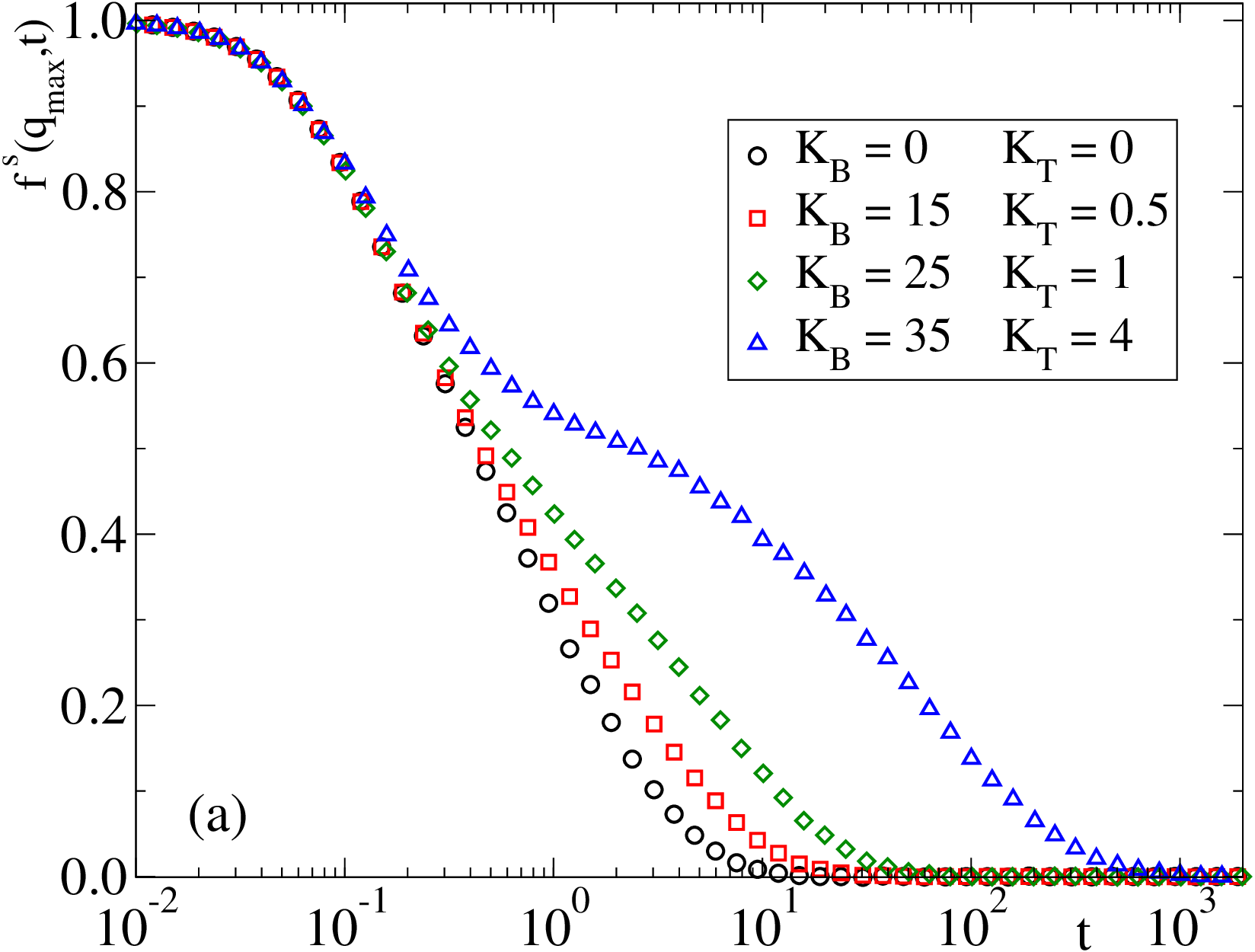}
\newline
\newline
\includegraphics[width=0.86\linewidth]{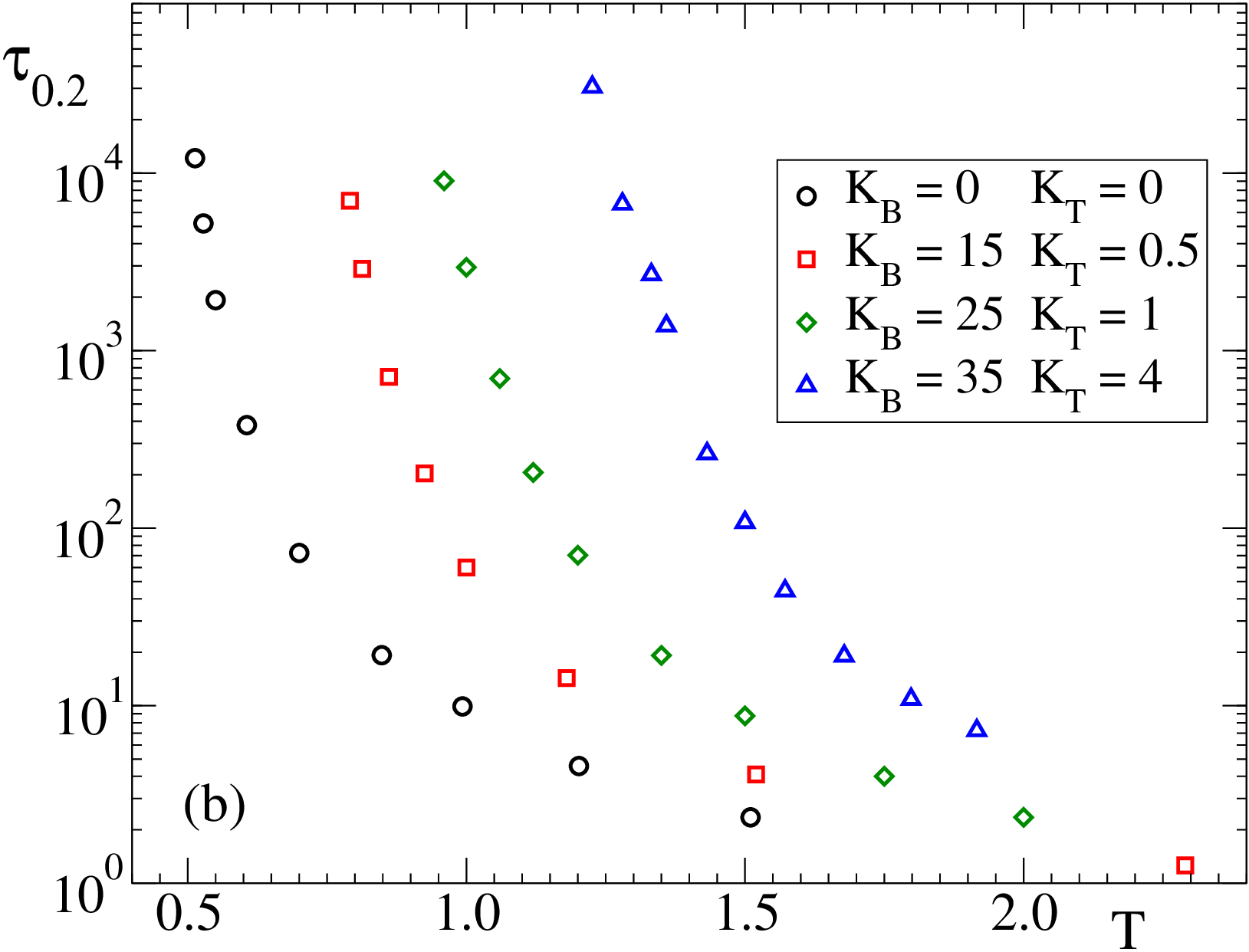}
\newline
\caption{Panel (a): self-density correlator $f^{\rm s}(q_{\rm max},t)$,
for several values of the barrier strength at fixed temperature $T=1.5$.
Panel (b): temperature dependence of the relaxation times $\tau_{0.2}$
of $f^{\rm s}(q_{\rm max},t)$, for the former values of the barrier strength.}
\label{fig:fsqttau02}
\end{center}
\end{figure}

Now we characterize dynamic features associated to the caging regime and the structural $\alpha$-relaxation.
Fig.~\ref{fig:fsqttau02}a shows the self-density correlator $f^{\rm s}(q,t)$ at fixed $T = 1.5$ 
and for several values of the barrier strength. The former is defined as 
$f^{\rm s}(q,t) = N^{-1}\langle \sum^{N}_{j=1}\exp[i {\bf q}\cdot ({\bf r}_j (t) - {\bf r}_j (0))]\rangle$. 
The sum is done over the coordinates ${\bf r}_j$ of all the $N$ monomers in the system.
In all the  cases the correlator is evaluated at the 
maximum, $q_{\rm max} \approx 7$ \cite{Bernabei:JCP2009}, of the static structure factor 
$S(q) = N^{-1}\langle \sum^{N}_{j,k=1}\exp[i {\bf q}\cdot ({\bf r}_j (0) - {\bf r}_k (0))]\rangle$.
We observe that increasing the strenght of the internal barriers at fixed $\rho$ and $T$ 
leads to slower dynamics. In the fully flexible case $f^{\rm s}(q,t)$ decays to zero in a single step.
On increasing the strength of the internal barriers $f^{\rm s}(q,t)$ exhibits 
the standard behavior in the proximity of a glass transition.
After the initial transient regime, $f^{\rm s}(q,t)$ shows a first decay to a plateau, which is associated to 
the caging regime, i.e., the temporary trapping of each particle by its neighbors.
At long times, a second decay is observed from the plateau to zero. 
This corresponds to the structural $\alpha$-relaxation. Similar trends are displayed
by the density-density correlator (not shown), defined as
$f(q,t) = \langle \rho({\bf q},t)\rho(-{\bf q},0)\rangle/\langle \rho({\bf q},0)\rho(-{\bf q},0)\rangle$,
with $\rho({\bf q},t)=\sum^N_{j=1} \exp[i{\bf q}\cdot{\bf r}_j(t)]$.

Let us define $\tau_{0.2}$ as the time for which $f^{\rm s}(q_{\rm max},\tau_{0.2}) = 0.2$,
and $\tau^{\rm K}_q$ as that obtained from fitting the $\alpha$-decay
to a Kohlrausch-Williams-Watts (KWW) function,
$A_q \exp[-(t/\tau^{\rm K}_q)^{\beta}]$ (with $A_q, \beta < 1$).
Both times correspond to a significant decay from the plateau, and therefore can be used as operational
definitions of the $\alpha$-relaxation time $\tau_{\alpha}$. 
Fig.~\ref{fig:fsqttau02}b shows $\tau_{0.2}$ as a funcion of $T$, 
for different values of the bending and torsional constants (results for $\tau^{\rm K}_q$ are analogous).
As observed in the analysis of the self-correlators, increasing the chain stiffness 
slows down the dynamics. At fixed temperature, the relaxation time for the stiffest 
investigated chains increases by several decades with respect to the fully-flexible case.

The dynamic trends displayed in Fig.~\ref{fig:fsqttau02}  
demonstrate that intramolecular barriers constitute
an additional mechanism for dynamic arrest, coexisting with the
general packing effects induced by density and temperature.
Now we discuss this scenario within the framework of the (ideal) MCT. 
We briefly summarize the basic concepts and predictions of the theory. 
Extensive reviews can be found, e.g., in 
Refs.~\cite{Gotze:RepProgPhys1992,Gotze:JPCM1999,Gotze:LesHouches1991,Reichman:JSTAT2005,Das:RevModPhys2004,Gotze:Book2009}. 
On approaching a glass transition from the ergodic phase, density fluctuations decay in a slower fashion,
remaining frozen in amorphous configurations when the glass transition occurs.
MCT describes this phenomenon  as a feedback mechanism driven by the slow density fluctuations.
By starting from the fundamental Liouville equation of motion and  
using the Mori-Zwanzig projection operator formalism, an integro-differential equation is obtained for the 
density-density correlator:
\begin{eqnarray}
\ddot{f}(q,t) + \frac{q^2k_BT}{mS(q)}  f(q,t) \hspace{2 cm} \nonumber \\
+\frac{q^2k_BT}{mS(q)}\int^{t}_{0}  dt' m(q,t-t') \dot{f}(q,t') = 0 .
\label{eq:fMCT}
\end{eqnarray}
The memory kernel
$m(q, t-t') \propto \langle R^{\rm f}_{\bf q} (0) R^{\rm f}_{\bf q} (t-t') \rangle$, 
is expressed in terms of the associated fluctuating forces  $R^{\rm f}_{\bf q}$ \cite{Reichman:JSTAT2005}.
In order to provide a closed solvable form of Eq. (\ref{eq:fMCT}), 
MCT introduces several approximations for the memory kernel.
These approximations are:
\\
\\
i)  The fluctuating force can be splitted in two terms: the regular (fast) contribution, 
linear in density fluctuations,
and a second term which can be expressed as a a linear combination of `mode pairs',
$\rho_{{\bf k}}\rho_{{\bf q-k}}$. The latter provides the slow contribution relevant
for the structural relaxation. Thus, the first approximation consists of neglecting the fast contribution.
\\
\\
ii) Convolution approximation: three-point static correlations are approximated as products
of static structure factors, 
\begin{eqnarray}
\langle \rho_{-{\bf q}}(0) \rho_{{\bf k}}(0) \rho_{{\bf q}-{\bf k}}(0) \rangle 
\approx N S(q)S(k)S(|{\bf q}-{\bf k}|) .
\label{eq:convol}
\end{eqnarray}
iii) Kawasaki approximation: dynamic four-point correlations are
factorized in terms of products of dynamic two-point correlations 
(see e.g., Ref. \cite{Reichman:JSTAT2005} for details).
Nowadays there is plenty of evidence that this approximation worsens on decreasing temperature, 
specially around the time scale of the $\alpha$-relaxation.
The breakdown of the former approximation is usually assigned
to the emergence of strong dynamic heterogeneities in the proximity of the glass transition
\cite{Glotzer:JNCS2000,Lacevic:JCP2003,Berthier:Science2005,Szamel:PRE2006,Berthier:JCP2007}.
\\
\\
After applying the former approximations,  the memory kernel 
$m(q,t)$ becomes  bilinear  in $f(q,t)$:
\begin{equation}
m(q,t) = \int\frac{d^3 {\bf k}}{(2\pi)^3}{\cal V}({\bf q},{\bf q}-{\bf k})f(k,t)f(|{\bf q}-{\bf k}| ,t) .
\label{eq:mMCT}
\end{equation}
The vertex ${\cal V}({\bf q},{\bf q}-{\bf k})$ is given by:
\begin{eqnarray}
{\cal V}({\bf q},{\bf q}-{\bf k}) = \frac{\rho}{2q^4} 
S(q)S(k)S(|{\bf q}-{\bf k}|) \times \nonumber \\
\left[{\bf q}\cdot{\bf k}c(k) + {\bf q}\cdot ({\bf q}-{\bf k})c(|{\bf q}-{\bf k}|)\right]^2 ,
\label{eq:vertMCT}
\end{eqnarray}
where $c(q)$ is the direct correlation function \cite{Hansen:Book1986}.
Eq.~(\ref{eq:fMCT}) constitutes a closed set of coupled equations which can be solved self-consistently,
provided $S(q)$ and $c(q)$ are known. The latter are {\it external} inputs in the MCT equations.
Since static correlators contained in the vertex vary with the control parameters 
(e.g., density, temperature, or barrier strength), the MCT equations (\ref{eq:mMCT}) establish
a direct connection between statics and dynamics. Moreover,
the former static correlators can be related to the interaction potential through
closure relations from liquid state theories \cite{Hansen:Book1986}.
With this, MCT provides a first-principle approach for the slow relaxation of density correlators.

Recently, Chong and co-workers have derived MCT equations for simple models of polymer melts
\cite{Chong:PRL2002,Chong:PRE2007}.
By exploiting the polymer reference interaction site model (PRISM) \cite{Schweizer:AdvChemPhys1997}, the MCT equations 
are considerable simplified.  This is achieved by replacing
site-specific intermolecular surroundings of a monomer by an averaged one 
(equivalent site approximation), whereas the full intramolecular dependence
is retained in the MCT equations \cite{Chong:PRL2002,Chong:PRE2007}.
The so-obtained scalar MCT equations of motion, memory kernel and vertex 
for polymer chains are formally {\it identical}
to Eqs. (\ref{eq:fMCT},\ref{eq:mMCT},\ref{eq:vertMCT}). 
The polymer character of the system only enters {\it implicitly} through
the PRISM relation \cite{Schweizer:AdvChemPhys1997} $\rho c(q) =1/\omega(q) -1/S(q)$, 
which differs from the Ornstein-Zernike
equation \cite{Hansen:Book1986}, $\rho c(q)= 1-S^{-1}(q)$, for monoatomic systems.
The quantity $\omega (q)$ is the chain form factor, defined as
%
%
%
\begin{equation} 
\omega (q) = \frac{1}{N_{\rm c}N_{\rm m}}\sum^{N_{\rm c}}_{I=1}\sum^{N_{\rm m}}_{a,b=1}
\left\langle\exp[i {\bf q}\cdot ({\bf r}^I_a (0) - {\bf r}^I_b (0))]\right\rangle,
\label{eq:wq}
\end{equation}
where ${\bf r}^I_a$ are the coordinates of the $a$th monomer in the $I$th chain. $N_{\rm c}$ 
is the total number of chains.
The use of the former MCT equations is a priori justified for polymers of variable stiffness.
Indeed, it has been shown that the PRISM  approximations
retain their validity not only in  the fully-flexible limit \cite{Aichele:PRE2004}, but also when
strong intramolecular barriers are present \cite{Bernabei:JCP2009}.

For the case of the self-density correlators $f^{\rm s}(q,t)$, the MCT equations 
are different from the monoatomic case. The former are obtained by summation of the
diagonal terms of the  self site-site density correlators. The latter are given by
\begin{equation} 
F^{\rm s}_{ab}(q,t) = \frac{1}{N_{\rm c}}\sum^{N_{\rm c}}_{I=1}
\left\langle\exp[i {\bf q}\cdot ({\bf r}^I_a (0) - {\bf r}^I_b (0))]\right\rangle,
\label{eq:Fsab}
\end{equation}
with indices defined as in Eq.~(\ref{eq:wq}). The correlators $F^{\rm s}_{ab}(q,t)$ are determined
by solving the corresponding MCT matrix equations (see Ref.~\cite{Chong:PRE2007}).

Ideal MCT predicts a sharp transition from an ergodic liquid to an arrested state (glass), 
at a given value of the relevant control parameter (temperature in the present case).
At the transition (or `critical') temperature $T = T_{\rm c}$, the non-ergodicity parameter, defined as 
$f_q = \lim_{t \rightarrow \infty} f(q,t)$, jumps from zero to a nonzero value $f_q^{\rm c}$.
The latter is called the {\it critical} non-ergodicity parameter. 
By taking the limit $t \rightarrow \infty$
in the MCT equations, one finds the relation


\begin{equation}
\frac{f_q}{1 -f_q}= 
\int \frac{d^3 {\bf k}}{(2\pi)^3}{\cal V}({\bf q},{\bf q}-{\bf k})f_{|{\bf q}-{\bf k}|}f_k .
\label{eq:nonergfac}
\end{equation}
Eq. (\ref{eq:nonergfac}) always has the trivial solution $\{f_q \}=0$.
Glassy states take place when solutions $f_q >0$ also exist.
The temperature at which the jump from zero to nonzero solutions occurs
defines $T_{\rm c}$. The corresponding solutions define
the critical non-ergodicity parameters.

The separation parameter, $\epsilon_T= (T -T_{\rm c})/T_{\rm c}$
measures the distance to the critical temperature. We are interested in the behavior 
of $f(q,t)$ in the ergodic fluid, i.e., for $\epsilon_T > 0$.
For small values of $\epsilon_T$, MCT predicts several
asymptotic laws for dynamic observables \cite{notebreak}, which are characterized by several dynamic exponents. 
Thus, the decay from
the plateau follows a von Schweidler expansion $f(q,t) = f_q^{\rm c}-h_q t^b +h_q^{(2)} t^{2b}+ O(t^{3b})$,
and diffusivities and $\alpha$-relaxation times obey $D^{-1}, \tau_{\alpha} \sim (T-T_{\rm c})^{-\gamma}$.
The exponents of these asymptotic laws are related to  
the so-called exponent parameter $\lambda$, which is the only independent one.
The MCT expression for $\lambda$ is determined by the static correlators
$S(q)$ and $c(q)$ evaluated at $T = T_{\rm c}$, and by the critical
non-ergodicity parameters $f_q^{\rm c}$ (see, e.g., \cite{Franosch:PRE1997,Bernabei:JCP2009}).

\begin{figure}
\begin{center}
\includegraphics[width=0.85\linewidth]{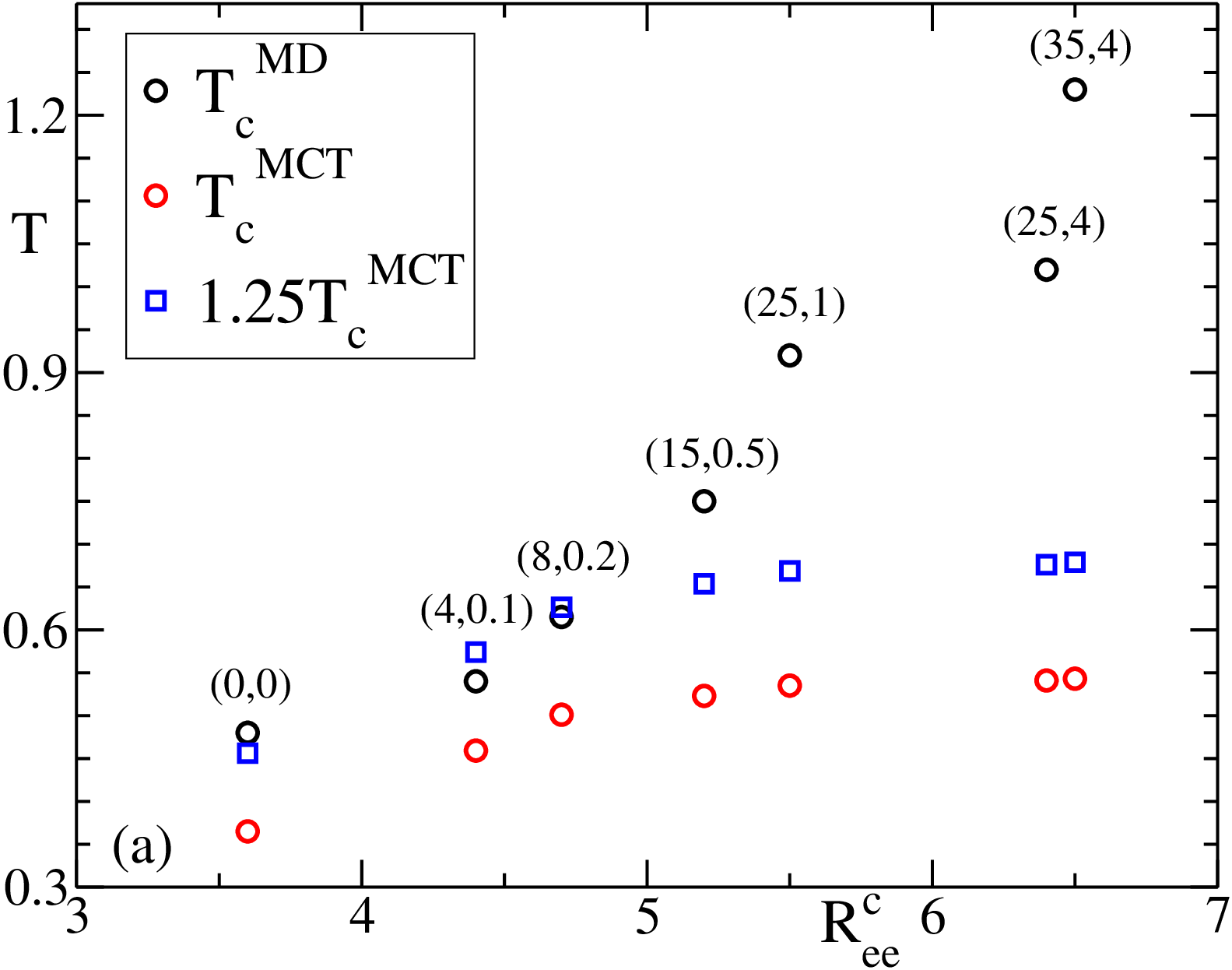}
\newline
\newline
\includegraphics[width=0.85\linewidth]{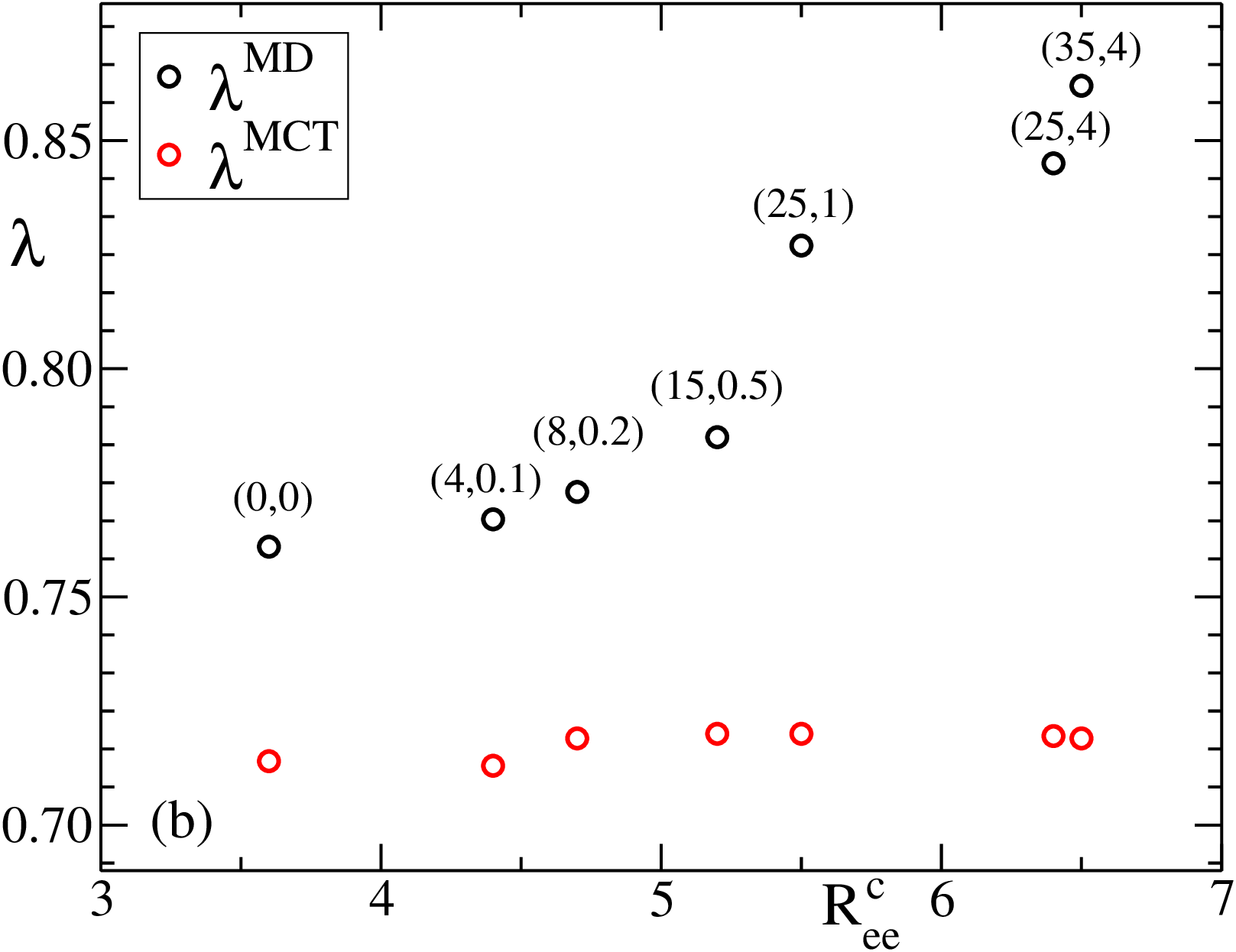}
\newline
\caption{Critical temperature $T_{\rm c}$ (a)  
and $\lambda$-exponent (b)  as a function of the end-to-end radius $R_{\rm ee}^{\rm c}$
(measured at $T_{\rm c}^{\rm MD}$).
The theoretical values $T_{\rm c}^{\rm MCT}$ and $\lambda^{\rm MCT}$  
are compared with the simulation values $T_{\rm c}^{\rm MD}$ and $\lambda^{\rm MD}$
(see text). The respective values of the bending and torsional strength $(K_{\rm B}, K_{\rm T})$
are indicated on top of each symbol for $T_{\rm c}^{\rm MD}$ in (a) and for $\lambda_{\rm c}^{\rm MD}$ in (b).}
\label{fig:lambdaTc}
\end{center}
\end{figure}

We solved Eqs.~(\ref{eq:fMCT}) and (\ref{eq:nonergfac}) by combining
simulation results of $\omega(q)$ with the PRISM equation  $\rho c(q) = 1/\omega(q) - 1/S(q)$  
and the Percus-Yevick closure relation \cite{Hansen:Book1986}. Details of the numerical procedure 
for solving (\ref{eq:nonergfac}) can be found
in Ref.~\cite{Bernabei:JCP2009}. Numerical integration of the density correlators
was performed following the method of Ref.~\cite{Gotze:JStaPhys1996}.
It often happens in the analysis of experiments or simulations
that numerical solutions of the MCT equations are not available.
In such cases, a phenomenological analysis can be performed, and the values of $T_{\rm c}$
and the former dynamic exponents  can be obtained as fit parameters 
from the  experimental or simulation data.
Consistency of the analysis requires that 
the exponents, which are obtained from independent fits to different scaling laws, 
are related to the same $\lambda$-parameter, as predicted by the theory.
This consistency test was done in the analysis of our simulation data
(see Refs.~\cite{Bernabei:PRL2008,Bernabei:JCP2009} for a detailed explanation), providing
different values of $T_{\rm c}$ and $\lambda$ for each barrier strength.
These values obtained {\it from simulations} can be compared with the values
provided by solution of the MCT equations.

This comparison is shown in Fig.~\ref{fig:lambdaTc}. Superscripts `MD' and `MCT'
are used respectively for simulation and theoretical values.
The data are represented as a function
of the end-to-end radius, which quantifies chain stiffness. 
A clear correlation between the barrier strength 
and the values of $T^{\rm MD}_{\rm c}$ and $\lambda^{\rm MD}$ is unambiguously demonstrated. 
The interplay between monomer packing effects and
intramolecular barriers \cite{Bernabei:JCP2009} induces a progressive increase 
of $T^{\rm MD}_{\rm c}$ at fixed density.
We note that from the fully-flexible limit $(K_{\rm B}, K_{\rm T}) = (0,0)$ to barriers
with $(K_{\rm B}, K_{\rm T}) = (15,0.5)$,
the data sets for $T^{\rm MD}_{\rm c}$ and $T_{\rm c}^{\rm MCT}$ 
roughly display the same slope.
As usual, there is a shift factor between simulation and theoretical temperatures
(here $T_{\rm c}^{\rm MD}/T_{\rm c}^{\rm MCT} \approx 1.25$), which may originate
from the mean-field character of MCT \cite{Gotze:Book2009}. 
The range of barrier strength for which $T_{\rm c}^{\rm MCT}$ and $T_{\rm c}^{\rm MD}$ 
are roughly parallel is significant. Indeed, for $(K_{\rm B} ,K_{\rm T}) = (8,0.2)$
the end-to-end radius $R_{\rm ee}^{\rm c}$ is a 30\% larger than for fully-flexible chains.
However, a strong discrepancy between simulation and theory becomes evident 
on increasing the barrier strength from $(K_{\rm B}, K_{\rm T}) = (15,0.5)$.
While beyond this point $T_{\rm c}^{\rm MCT}$ seems to approach an asymptotic limit,
$T_{\rm c}^{\rm MD}$ increases up to  1.23 for the stiffest investigated chains.

\begin{figure}
\begin{center}
\includegraphics[width=0.87\linewidth]{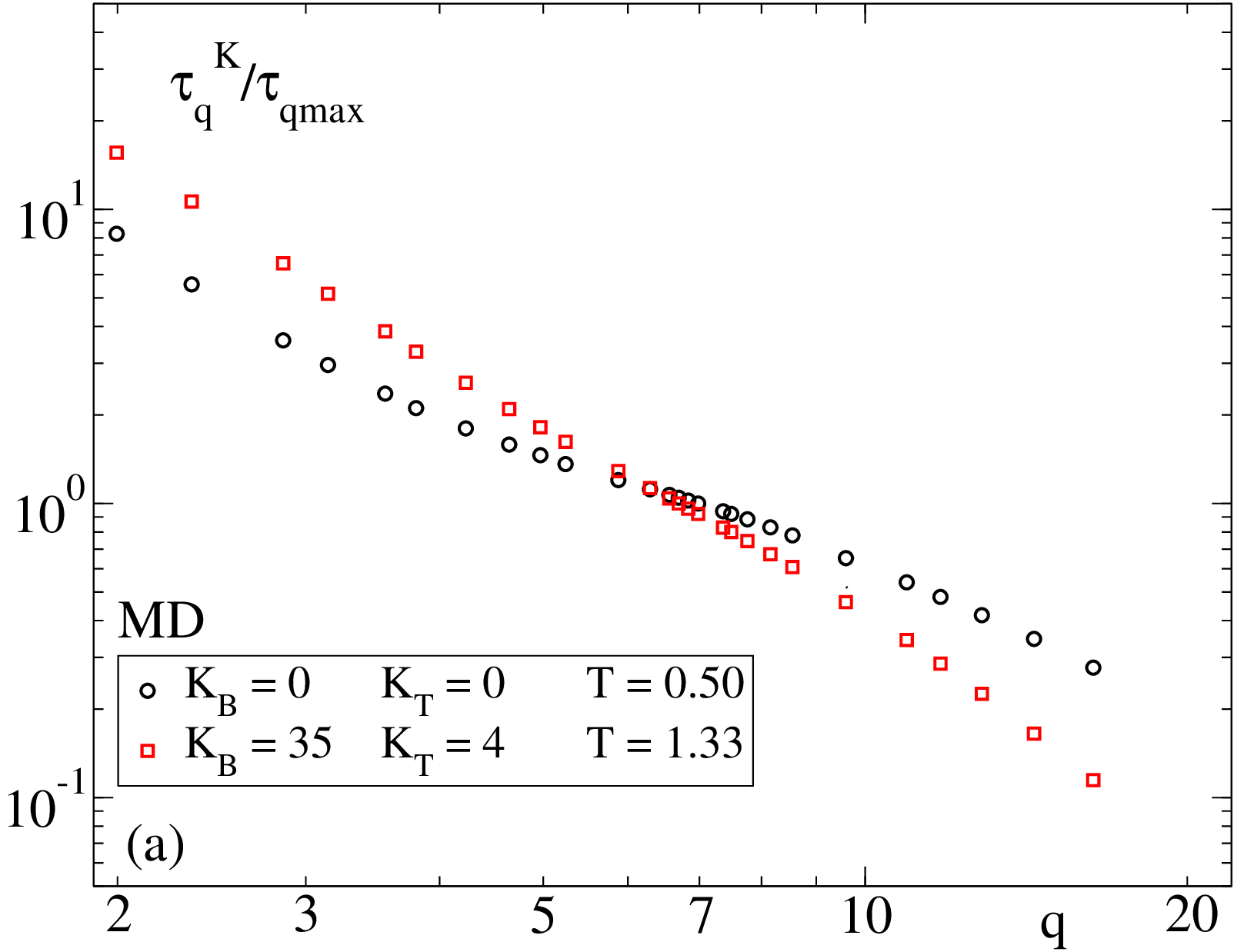}
\newline
\newline
\includegraphics[width=0.87\linewidth]{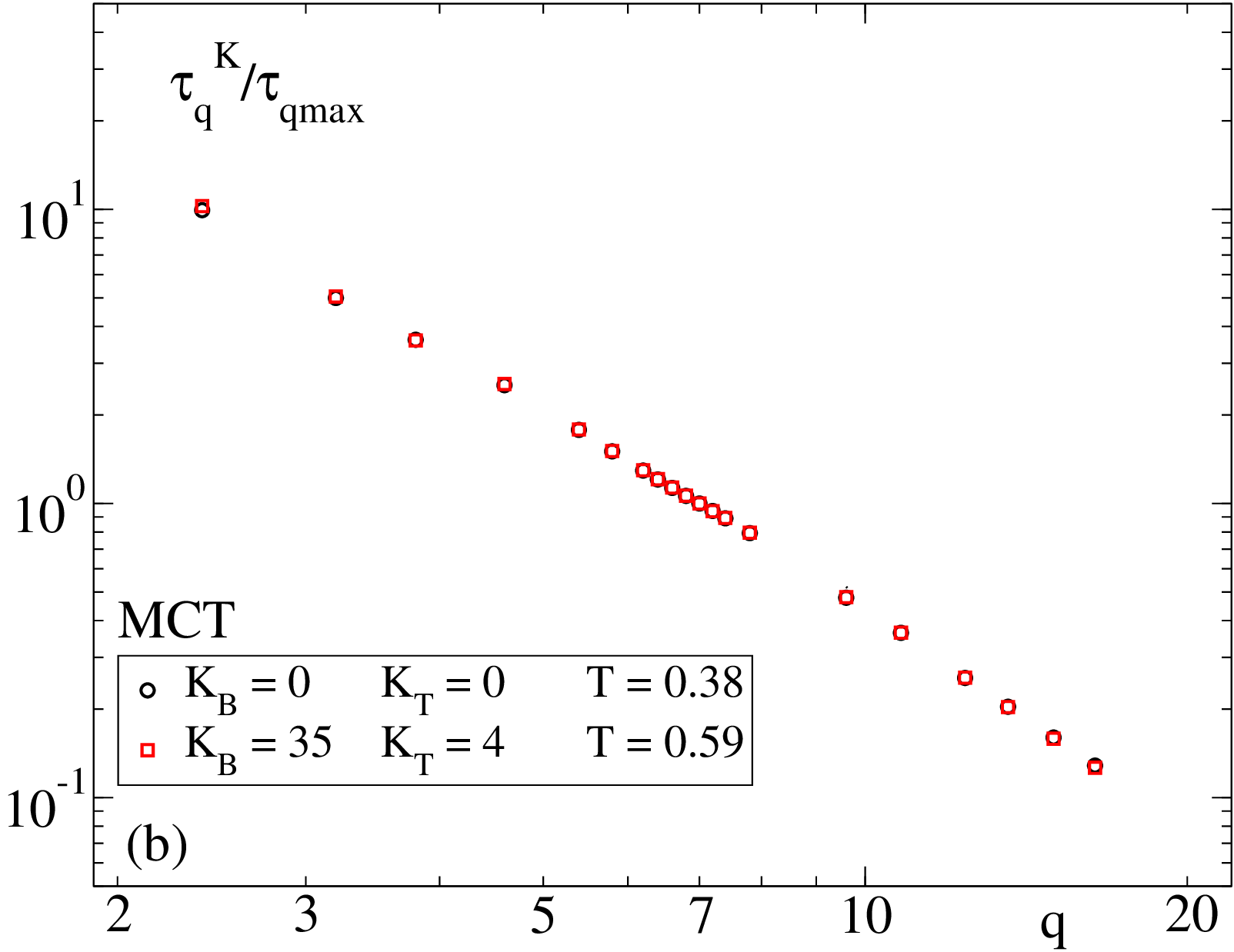}
\newline
\caption{For fully-flexible and stiffest investigated  chains,
$q$-dependence of the KWW times for self-density correlators $f^{\rm s}(q,t)$.
Data are rescaled by the respective KWW times at $q_{\rm max}$. Data in panels (a) and (b)
correspond respectively to simulation results and MCT solutions at
selected temperatures (see legend). }
\label{fig:tauq2}
\end{center}
\end{figure}

Similar trends are observed for the $\lambda$-exponent. Simulation values increase from 
$\lambda^{\rm MD} = 0.76$ for fully-flexible chains to $\lambda^{\rm MD} = 0.86$ 
for the stiffest investigated chains. The first ones are
typical of simple glass-formers as the archetype hard-sphere fluid
($\lambda = 0.74$ \cite{Franosch:PRE1997}), where dynamic arrest is driven by packing effects. 
The largest ones are similar to those observed in realistic models 
of polymer melts which incorporate the full chemical structure of the chains
\cite{Colmenero:JPCM2007,Narros:Thesis,Capponi:JCP2009}. 
On the contrary, the theoretical exponent exhibits
a very weak variation, $0.71 \le \lambda^{\rm MCT} \le 0.72 $, over the investigated range
of barrier strength.

These discrepancies in the case of strong intramolecular barriers 
are also reflected in the $q$-dependence of density correlators
computed from simulations and from solution of the MCT equations.
In both cases we fitted the corresponding $\alpha$-decay to a KWW function (see above).
Fig. \ref{fig:tauq2} compares the $q$-dependence, at fixed $T$, 
of the KWW time $\tau^{\rm K}_q$ for the self-correlators $f^{\rm s}(q,t)$,
as obtained from simulations and from theory. 
Results are presented for the fully-flexible case and for the stiffest investigated chains.

Before discussing such results, some points must be clarified.
As mentioned above, the mean-field character of MCT usually yields a temperature shift between
simulation  and theory (see Fig.~\ref{fig:lambdaTc}a). 
Moreover, MCT times are affected by an undetermined constant factor \cite{Gotze:Book2009}.
Thus, a proper comparison between theory and simulation for time-dependent correlators 
can be done by rescaling $t$ by some characteristic relaxation time, 
and using a common separation parameter $\epsilon_{T}$ \cite{Chong:PRE2007}.
This is the case for the data of Fig.~\ref{fig:tauq2}. Thus, each data set is rescaled by the respective
KWW time $\tau_{q_{\rm max}}$ corresponding to $f^{\rm s}(q_{\rm max},t)$.
Temperatures of both panels correspond to $\epsilon_T \approx 0.04$ and 0.08
for respectively fully-flexible and stiff chains. In the rest of the article
we will present several comparisons between simulation data and MCT solutions.
It will be understood that the times and temperatures of the compared data obey the former criteria.

\begin{figure}
\begin{center}
\includegraphics[width=0.92\linewidth]{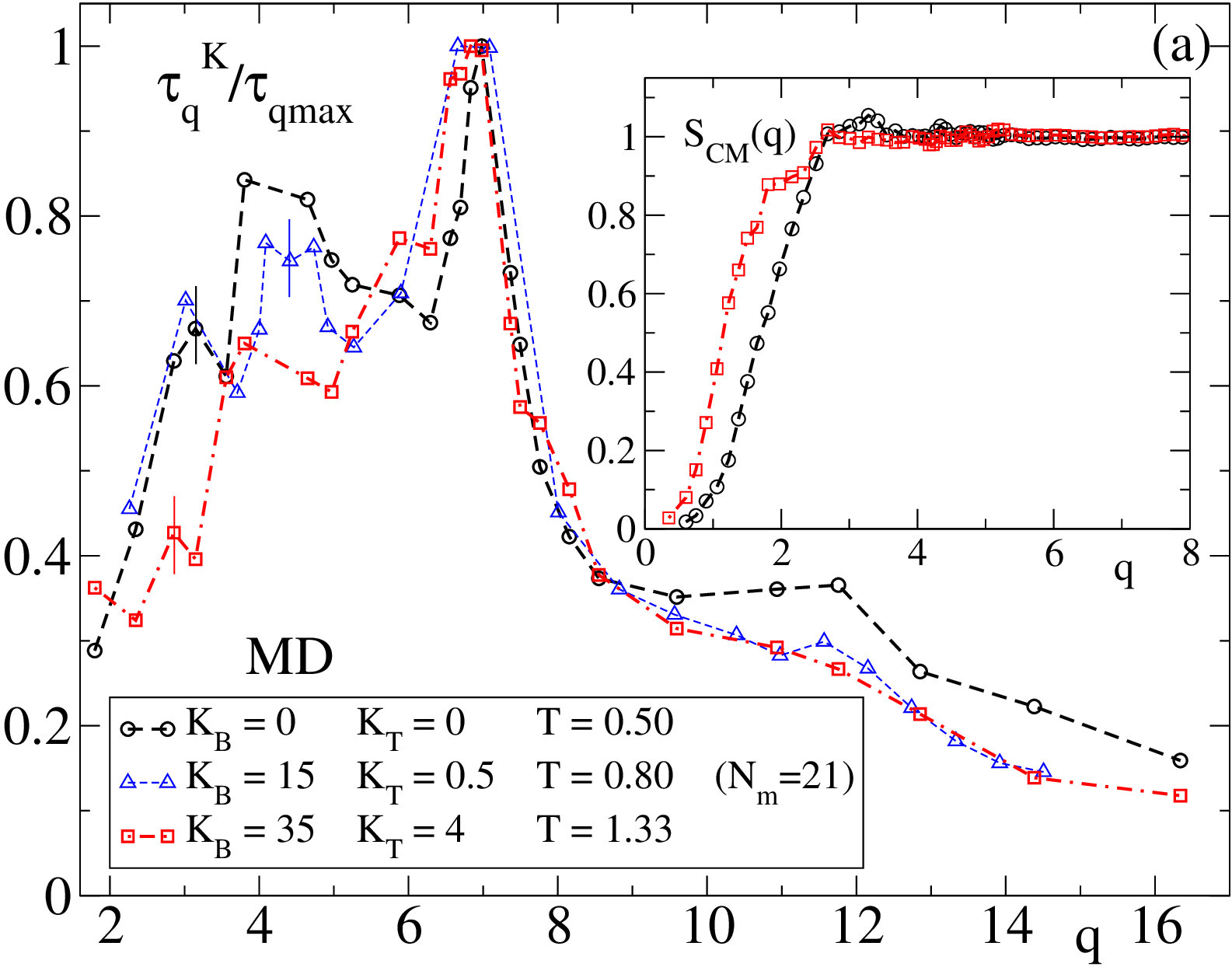}
\newline
\newline
\includegraphics[width=0.92\linewidth]{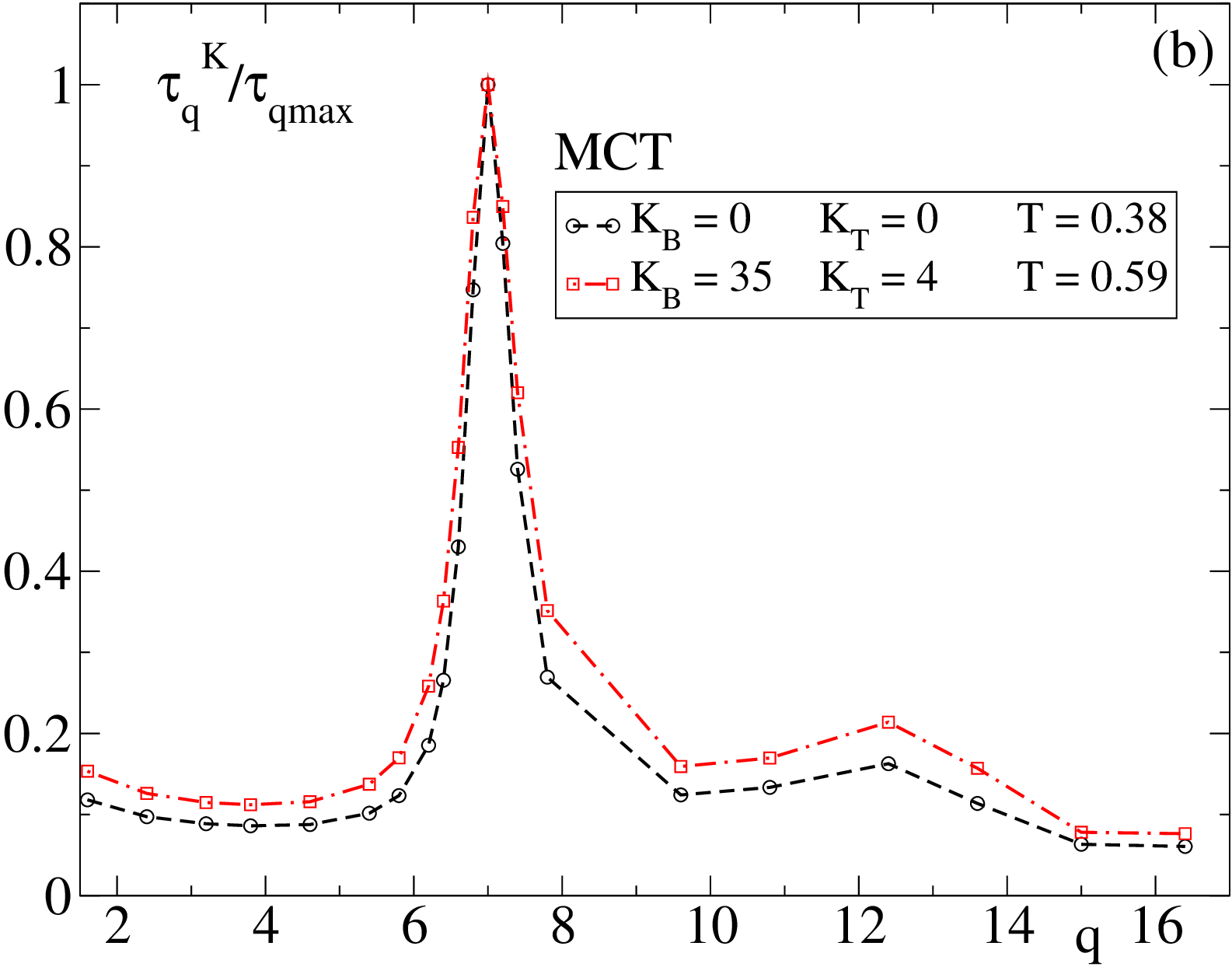}
\newline
\caption{Main panels: for fully-flexible and stiffest investigated  chains,
$q$-dependence of the KWW times for density-density correlators $f(q,t)$.
Data are rescaled by the respective KWW times at $q_{\rm max}$. Data in (a) and (b)
correspond respectively to simulation results and MCT solutions at
selected temperatures (see legend). Typical error bars are included in (a).
Dashed lines are guides for the eyes. Inset of panel (a):
Static structure factors for the chains centers-of-mass, $S_{\rm CM}(q)$.}
\label{fig:tauq}
\end{center}
\end{figure}

A clear disagreement between simulation and theoretical
trends becomes evident in Fig.~\ref{fig:tauq2}. 
The two sets of KWW times obtained from simulation show a rather different $q$-dependence,
which is more pronounced for the stiff chains. On the contrary, after rescaling by $\tau_{q_{\rm max}}$,
the theoretical sets become esentially identical. 
Fig \ref{fig:tauq} shows a similar comparison between simulation and theory for the rescaled KWW times
of the density-density correlators, $f(q,t)$.
For $q  \ge q_{\rm max}$ the theory reproduces qualitatively the shape of the relaxation times, 
which are modulated by the respective static structure factor $S(q)$ (not shown).
However, MCT fails at reproducing the broad peak at intermediate $q_{\rm C} \approx 4$ 
which is present in the simulation data.
This failure  was already noted for fully-flexible chains in Ref.~\cite{Chong:PRE2007},
and is confirmed here 
for the general case with intramolecular barriers. 
Apparently (note the error bars), the peak does not shift significantly
and decreases its intensity as chains become stiffer, leading to a shoulder.
In previous works \cite{Aichele:EPJE2001b} on similar fully-flexible bead-spring chains of $N_{\rm m} = 10$, the
value of $q_{\rm C}$ has been identified with $2\pi/R_{\rm g}$, where $R_{\rm g}$
is the chain radius of gyration.
Data of Fig.~\ref{fig:tauq}a do not seem compatible with this assignment. 
Appart from results for fully-flexible and stiffest investigated chains of $N_{\rm m} =10$, 
we include data for $(K_{\rm B}, K_{\rm T})= (15, 0.5)$ of additional simulations with $N_{\rm m} =21$. 
With this, the data sets of Fig.~\ref{fig:tauq}a cover a significant variation in $R_{\rm g}$.
Namely, for  $(K_{\rm B}, K_{\rm T})= (0, 0)$, (35,4), and (15,0.5)  
we respectively find $2\pi/R_{\rm g} =4.2$, 2.8, and 2.0.
Thus, the observation $q_{\rm C} \approx 2\pi/R_{\rm g}$ for fully-flexible chains is apparently fortuitous.
The associated length scale $2\pi/q_{\rm C} \approx 1.6\sigma$ rather seems to be a characteristic
feature which does not depend significantly on the barrier strength.  
We will come back to these points in Section IVc.


\begin{center}
{\bf B. Chain dynamics}
\end{center}

In this subsection we compare simulation and theoretical results for the dynamics of the Rouse modes.
First we briefly summarize the assumptions and main predictions of the Rouse model.
The starting point is a tagged gaussian
chain of $N_{\rm m}$ monomers  connected by
harmonic springs of constant $3k_{\rm B}T/b^2$, with
$b$ the bond length. The effective interaction experienced by the monomers  is
given by a friction coefficient $\zeta$ and a set of stochastic forces ${\bf f}_j$.
Excluded volume interactions are neglected.
The chain motion is mapped onto a set of $N_{\rm m}$
normal modes (Rouse modes) labelled by $p=0,1,2,...,N_{\rm m}-1$, of wavelength $N/p$, and defined as
\cite{DoiEdwards:Book1986,notenorm}
${\bf X}_p(t) =  \sum_{j=1}^{N_{\rm m}}P_{jp}{\bf r}_j(t)$, 
with $P_{jp} = \sqrt{(2-\delta_{p0})/N_{\rm m}}\cos[(j-1/2)p\pi/N_{\rm m}]$. 
The chain center-of-mass coincides with ${\bf X}_0(t)/\sqrt{N_{\rm m}}$.
The mode correlators are defined as
$C_{pq}(t)=[\langle{\bf X}_p(0)\cdot{\bf X}_q(t)\rangle-\delta_{0,p\times q}\langle{\bf X}_p(0)\cdot{\bf X}_q(0)\rangle]/3N_{\rm m}$. 
For $p,q >0$ we define the matrix $\hat{C}_{pq}(0) \equiv  C_{pq}(0)$. 
In the Rouse model  the stochastic forces are fully spatial and time uncorrelated, i.e., 
$\langle {\bf f}_j(t) \cdot{\bf f}_k(t') \rangle = 6\zeta k_{\rm B}T\delta_{jk}\delta(t-t')$.
These properties of the random forces  lead to orthogonality and exponentiality  
of the Rouse modes \cite{DoiEdwards:Book1986}. Thus, the mode correlators obey
$C_{pq}(t) = \hat{C}_{pq}(0)\exp[-t/\tau_p]$, with 
$\hat{C}_{pq}(0) = \delta_{pq}(b^2/24N_{\rm m}^2)\sin^{-2}[p\pi/2N_{\rm m}]$
and $\tau_p=(\zeta b^2/12k_{\rm B}T) \sin^{-2}[p\pi/2N_{\rm m}]$. Accordingly,
for $p \ll N_{\rm m}$ the quantities $\hat{C}_{pp}(0)$ and $\tau_{p}$ scale as $\sim p^{-2}$.

\begin{figure}
\includegraphics[width=0.92\linewidth]{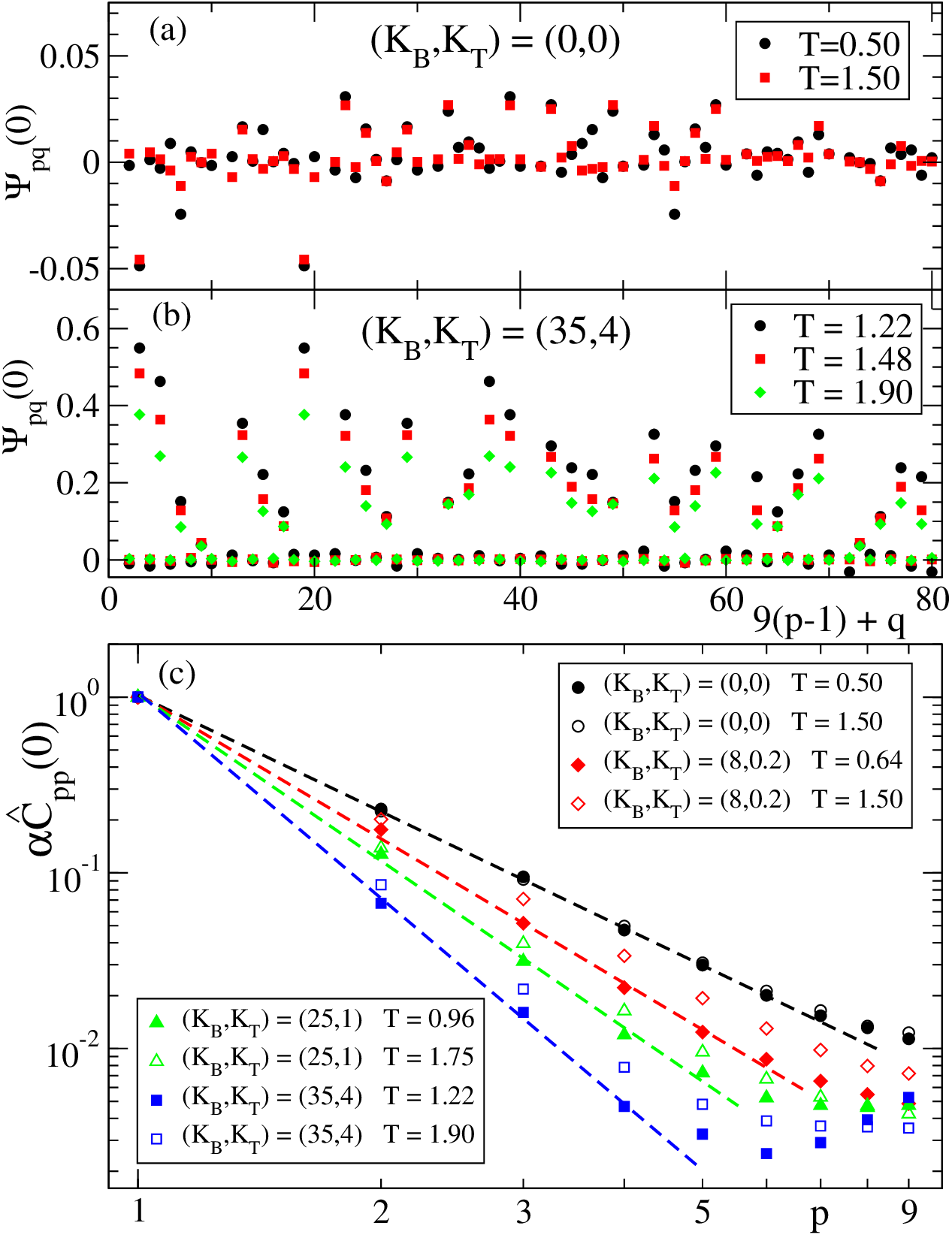}
\caption{Static intrachain correlations computed from simulations.
For each value of $(K_{\rm B}, K_{\rm T})$ (see legends) results include data at the highest
and lowest investigated $T$.
Panels (a) and (b): off-diagonal terms of $\Psi_{pq}(0)$ (see text) versus the combined variable $9(p-1)+q$.
Data in (a) and (b) respectively correspond to fully-flexible chains
and to the stiffest investigated chains.
Panel (c): diagonal terms $\hat{C}_{pp}(0)$ (see text).
Each data set corresponds to a fixed value of $(K_{\rm B}, K_{\rm T})$ and $T$ (see legend).
For clarity, each set has been rescaled by a factor $\alpha$ in order to get $\alpha \hat{C}_{11}(0) =1$ in all cases.
Dashed lines indicate approximate power-law behavior $\sim p^{-x}$. From top to bottom, $x = 2.2$, 2.7, 3.1, and 3.8.} 
\label{fig:rousestatic}
\end{figure}

In the following we show how the former scaling properties are strongly altered by the
introduction of intramolecular barriers. This is demonstrated
in Fig.~\ref{fig:rousestatic} for the case of  intrachain static correlations.
We show the off-diagonal terms 
of $\Psi_{pq}(0)=\langle {\bf X}_p(0) \cdot {\bf X}_q(0) /(X_p(0)X_q(0))\rangle$
(the diagonal terms are trivially $\Psi_{pp}(0) \equiv 1$). 
Data for fully-flexible chains  exhibit small
deviations from orthogonality, indeed $|\Psi_{pq}(0)| < 0.05$ for all $p\neq q$, independently
of $T$. Instead, orthogonality is clearly violated 
for strong intramolecular barriers.
Off-diagonal terms can take values  of even a  60\% percent of the diagonal ones.
Moreover, deviations are enhanced by decreasing temperature.
Fig.~\ref{fig:rousestatic}c shows results for the unnormalized diagonal terms  $\hat{C}_{pp}(0)$
(see above). In the low $p$-range the data can be described by an effective power law 
$\hat{C}_{pp}(0) \sim p^{-x}$. For fully-flexible chains we find approximate 
gaussian  behavior, $\hat{C}_{pp}(0) \sim p^{-2.2}$ \cite{Bennemann:EPJB1999}.
However, the introduction of internal barriers leads to strong non-gaussian behavior. 
On increasing the barrier strength, the effective exponent $x$ increases up to a value
of 3.8 for the stiffest case, $(K_{\rm B},K_{\rm T})= (35,4)$, at the lowest $T$. 
The most local effects of the intramolecular barriers 
are manifested by flattening  of  $\hat{C}_{pp}(0)$  at large $p$.

\begin{figure}
\includegraphics[width=0.85\linewidth]{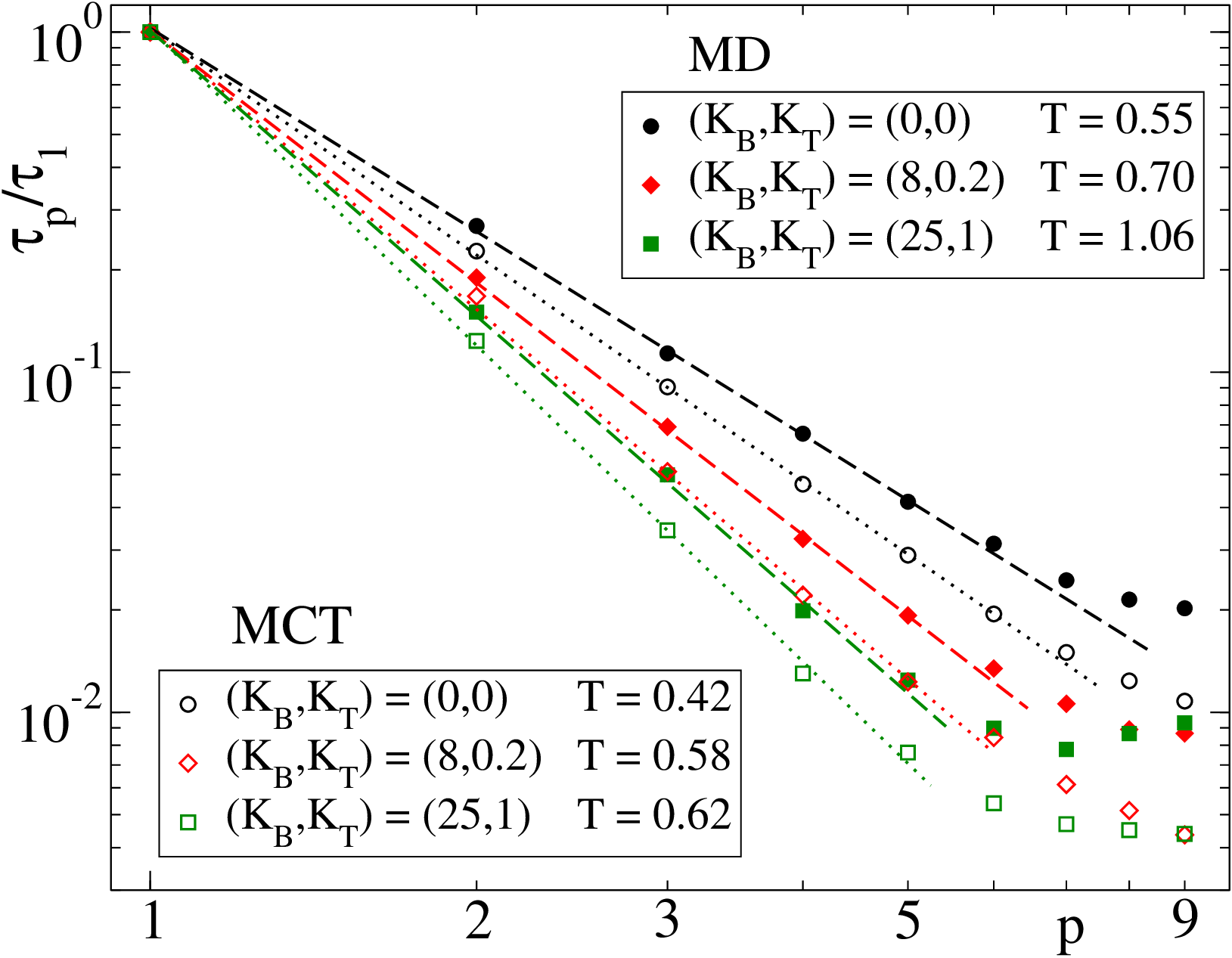} 
\caption{Simulation results (filled symbols) and MCT solutions (empty symbols),
for the $p$-dependence of the relaxation times $\tau_p$ of the 
mode correlators.
Each data set corresponds to a fixed value of $(K_{\rm B}, K_{\rm T})$ and $T$ (see legends).
For clarity, each set is rescaled by its respective $\tau_1$.
Dashed and  dotted lines indicate approximate power-law behavior $\sim p^{-x}$.
From top to bottom, simulations (dashed): $x = 2.0$, 2.4, 2.8; MCT (dotted): $x = 2.2$, 2.7, 3.1.} 
\vspace{-1 mm}
\label{fig:taup}
\end{figure}

The trends observed for intrachain static correlations have their dynamic counterparts.
Fig.~\ref{fig:taup} shows the relaxation times $\tau_p$, of the normalized mode correlators
$\Phi_{pp}(t) =  C_{pp}(t)/\hat{C}_{pp}(0)$, as a function of the mode index $p$. 
We display data for several values of the bending
and torsion constants $(K_B,K_T)$ and temperatures $T$.
The relaxation times have been operationally defined as $\Phi_{pp}(\tau_p) = 0.3$.
Data can be again described at low-$p$ by an effective power-law $\tau_p \sim p^{-x}$.
The observed trends are analogous to those found for the static correlations 
(Fig.~\ref{fig:rousestatic}c). Rouse behavior ($x =2$) is observed
only in the fully-flexible limit.
Again, as for the static amplitudes $\hat{C}_{pp}(0)$, $x$ is weakly dependent on $T$ \cite{notex}
but  strongly dependent on the barrier strength, taking higher values for stiffer chains.
The $x$-values for $\hat{C}_{pp}(0)$ and $\tau_p$ at the same 
$(K_{\rm B},K_{\rm T})$ and $T$ are  similar. This suggests that the structural origin
of the observed dynamic anomalies 
is mainly controlled by intrachain static correlations.

\begin{figure}
\begin{center}
\includegraphics[width=0.90\linewidth]{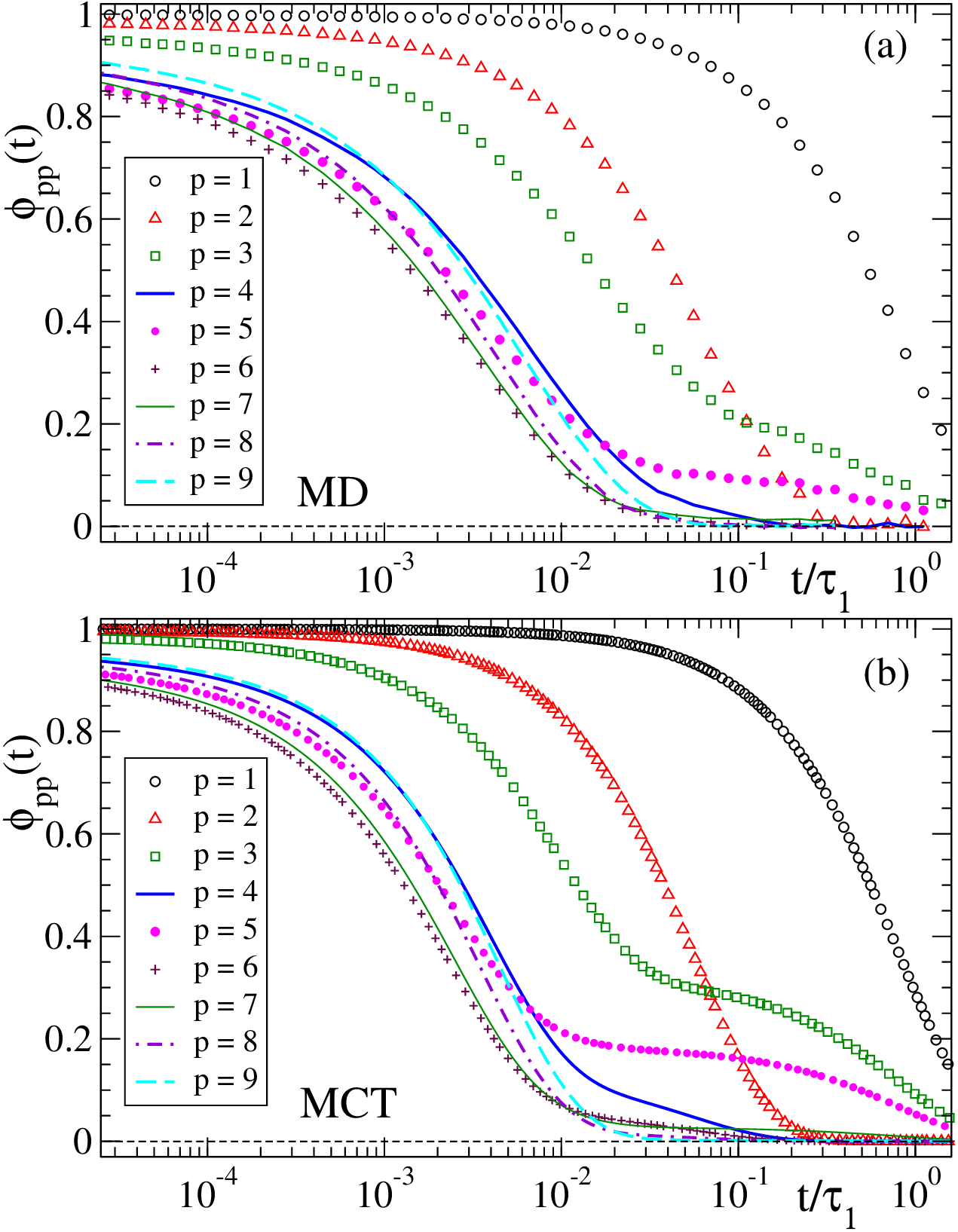}
\caption{Normalized mode correlators $\Phi_{pp}(t)$ for stiff chains with $(K_{\rm B},K_{\rm T}) = (35,4)$.
Panel (a): simulation results at $T = 1.48$. Panel (b): MCT numerical solutions at $T = 0.63$.
In both panels,  the absolute time is rescaled by the relaxation time $\tau_1$ of the $p=1$  mode.} 
\label{fig:rousemodes}
\vspace{-1 mm}
\end{center}
\end{figure}

Fig.~\ref{fig:rousemodes}a shows simulation results for the normalized mode correlators $\Phi_{pp}(t)$,
for $(K_{\rm B},K_{\rm T})=(35,4)$, at $T = 1.48$. Times are rescaled by the relaxation time
of the first mode, $\tau_1$.
Several salient features are revealed. First, the unambiguous presence of a long-time plateau 
for the modes $p=3$ and $p=5$, followed by an ultimate slow decay. 
It must be stressed that this feature is not related to the structural $\alpha$-relaxation.
Indeed, the plateau arises at times far beyond the $\alpha$-time scale ($\tau_{\alpha} \sim 5\times10^{-3}\tau_1$ 
for the considered $T$). This feature is instead intimately connected to the 
relaxation of the internal torsional degrees of freedom of the chain. 
Indeed we observe (not shown) that for fixed bending constant $K_{\rm B}$, the long-time plateau
tends to vanish as the value of the torsional constant $K_{\rm T}$ is decreased.

The observed long-time plateau constitutes a clear breakdown of the Rouse model, 
which predicts single, purely exponential decays of the mode correlators (see above). 
Its origin can be temptatively understood as follows. 
The relaxation of the $p$th-mode is equivalent to the relaxation of a harmonic
oscillation of wavelength $N/p$ . In the case of strong torsional barriers, 
the wavelengths of some particular modes probe characteristic lengths over which chain deformation
involves a strong energetic penalty (due to the presence of the barriers). 
Thus, at the time scales for which the barrier amplitudes are probed, the relaxation of such modes 
becomes strongly hindered, leading to the observed long-time plateau regime and ultimate slow relaxation. 
Another intriguing feature of  Fig.~\ref{fig:rousemodes}a, 
also inconsistent with the Rouse model, is the non-monotonous $p$-dependence 
of the mode correlators at intermediate times  prior to the long-time plateau (see data for $p > 4$).

Now we demonstrate that all the former dynamic features
can be rationalized in terms of the PRISM-based MCT approach
of Chong {\it et al.}. As exposed in Refs.~\cite{Chong:PRL2002,Chong:PRE2007}, the MCT equations 
for the unnormalized Rouse correlators 
$C_{pq}(t)$ are derived as the $q \rightarrow 0$
limit of the equations for the self site-site density correlators $F^{\rm s}_{ij}(q,t)$. 
The equations for $C_{pq}(t)$ read  \cite{Chong:PRL2002,Chong:PRE2007}:
\begin{eqnarray}
\ddot{C}_{pq}(t) + \frac{k_{\rm B}T}{mN_{\rm m}}\delta_{0p}\delta_{0q}+\frac{k_{\rm B}T}{m}\sum_{k=0}^{N_{\rm m}-1}E_{pk}C_{kq}(t) + \nonumber \\
\frac{k_{\rm B}T}{m}\sum_{k=0}^{N_{\rm m}-1}\int_{0}^{t}dt'm_{pk}(t-t') \dot{C}_{kq}(t')=0,
\label{eq:rouseMCT}
\end{eqnarray}
with $\hat{C}^{-1}_{pq}(0)$  the inverse
matrix of $\hat{C}_{pq}(0)$, and $E_{pq}=(1-\delta_{0,p\times q})\hat{C}^{-1}_{pq}(0)/N_{\rm m}$. 

The memory kernel is given by
$m_{pq}(t)=(\rho/6\pi^2)\int dk k^4 S(k)c^2(k)\sum_{i,j=1}^{N_{\rm m}}P_{ip}F^{\rm s}_{ij}(k,t)P_{jq} f(k,t)$
\cite{notemarkov,Allegra:AdvChemPhys1989,Harnau:JCP1995}.
Thus, prior to solve Eq.~(\ref{eq:rouseMCT}), we obtained the density-density correlators $f(k,t)$ 
and self site-site density correlators $F^{\rm s}_{ij}(k,t)$ 
from their respective MCT equations (see Ref. \cite{Chong:PRE2007}).
The static quantities $\hat{C}_{pq}(0)$ and $\hat{C}_{pq}^{-1}(0)$, 
which also enter  Eq.~(\ref{eq:rouseMCT}) as external inputs, were directly 
computed from the simulations at the respective lowest  investigated temperature. 

\begin{figure}
\begin{center}
\includegraphics[width=0.84\linewidth]{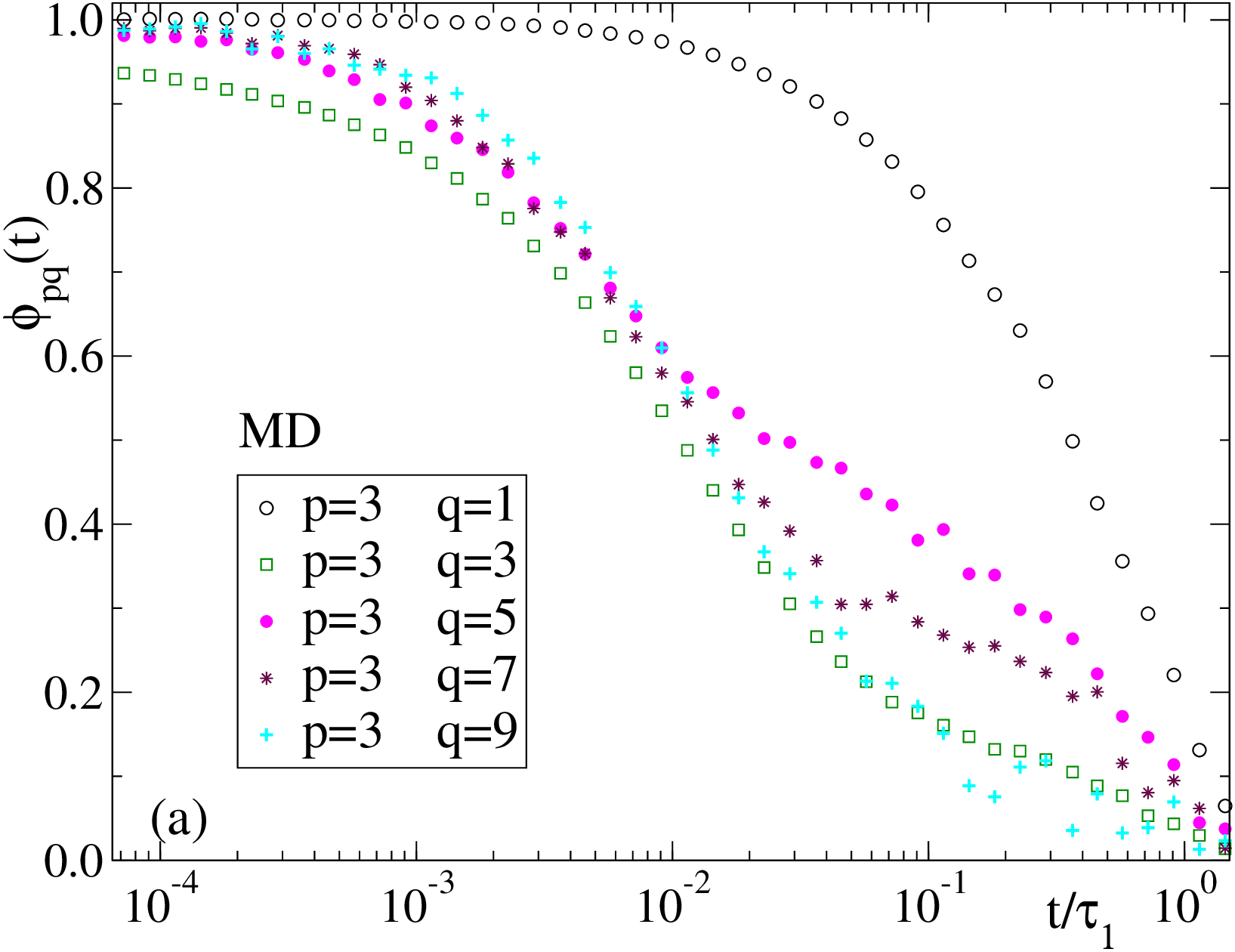}
\newline
\newline
\includegraphics[width=0.84\linewidth]{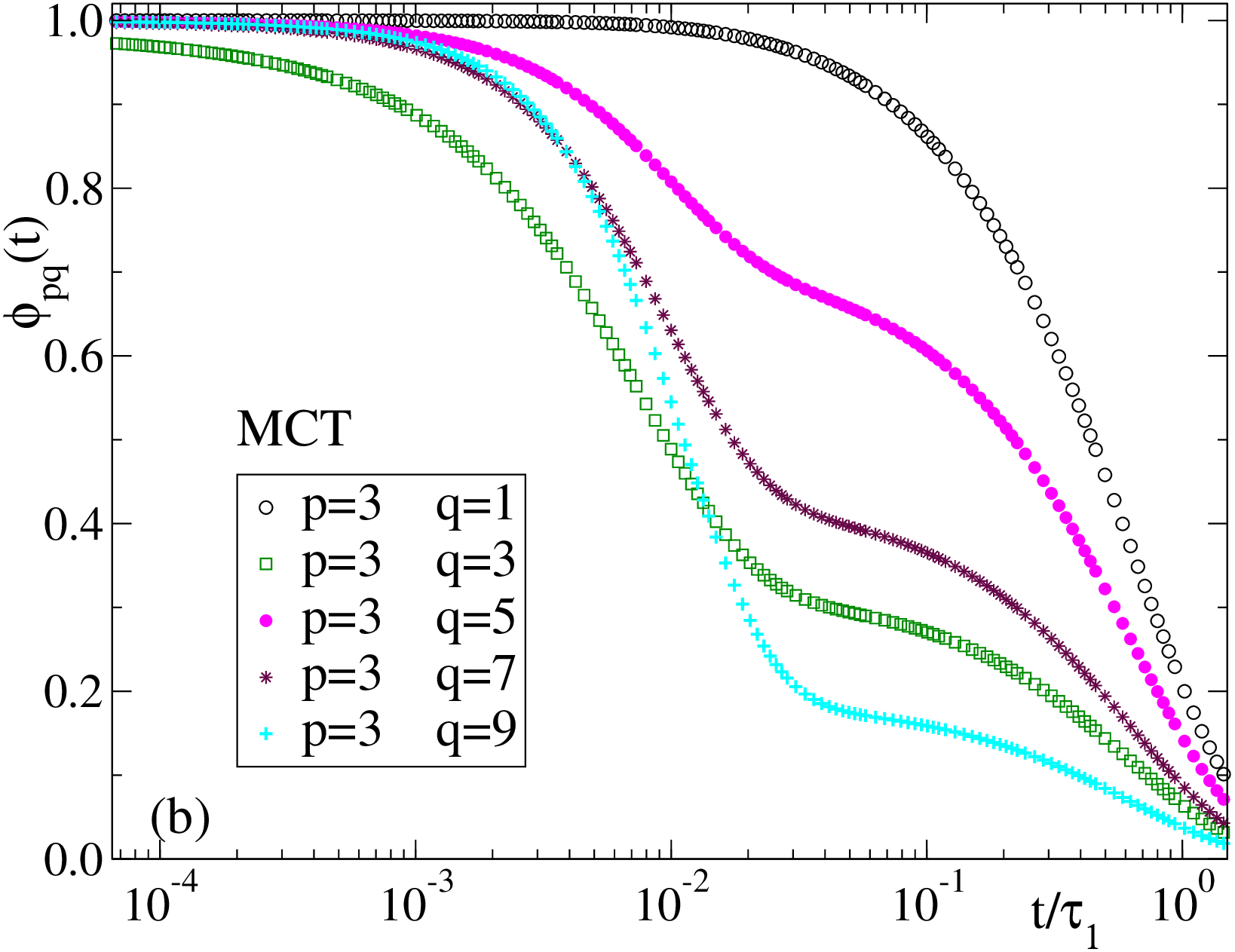}
\newline
\caption{Normalized cross-correlators $\Phi_{pq}(t)$ (for fixed $p=3$) 
of stiff chains with $(K_{\rm B},K_{\rm T}) = (35,4)$.
Panel (a): simulation results at $T = 1.48$. Panel (b): MCT numerical solutions at $T = 0.63$.
In both panels,  the absolute time is rescaled by the relaxation time $\tau_1$ of the $p=1$  mode.}
\label{fig:rousecross}
\end{center}
\end{figure}

Fig.~\ref{fig:rousemodes} shows a comparison, at $\epsilon_T \approx 0.2$,
of the MCT solutions for the normalized mode correlators $\Phi_{pp}(t)$ [panel (b)]
of the stiffest investigated chains, with the respective simulation results previously discussed
[panel (a)].  A full correspondence between MCT solutions and  simulation trends is obtained.
These include the long-time plateaux for $p=3$ and $p=5$, as well as the sequence
in the complex, non-monotonous $p$-dependence for $p>4$ at intermediate times.
As previously done for the simulation data, we can obtain the theoretical relaxation times $\tau_p$ 
from the condition $\Phi_{pp}(\tau_p) = 0.3$ in the theoretical correlators.
The $p$-dependence of the simulation and theoretical times are compared in Fig.~\ref{fig:taup},
at common $\epsilon_T \approx 0.2$, for several values of $(K_{\rm B},K_{\rm T})$.
Again, MCT solutions are in semiquantitative agreement with the anomalous trends of simulations,
with similar exponents for the effective power-laws.

\begin{figure}
\begin{center}
\includegraphics[width=0.84\linewidth]{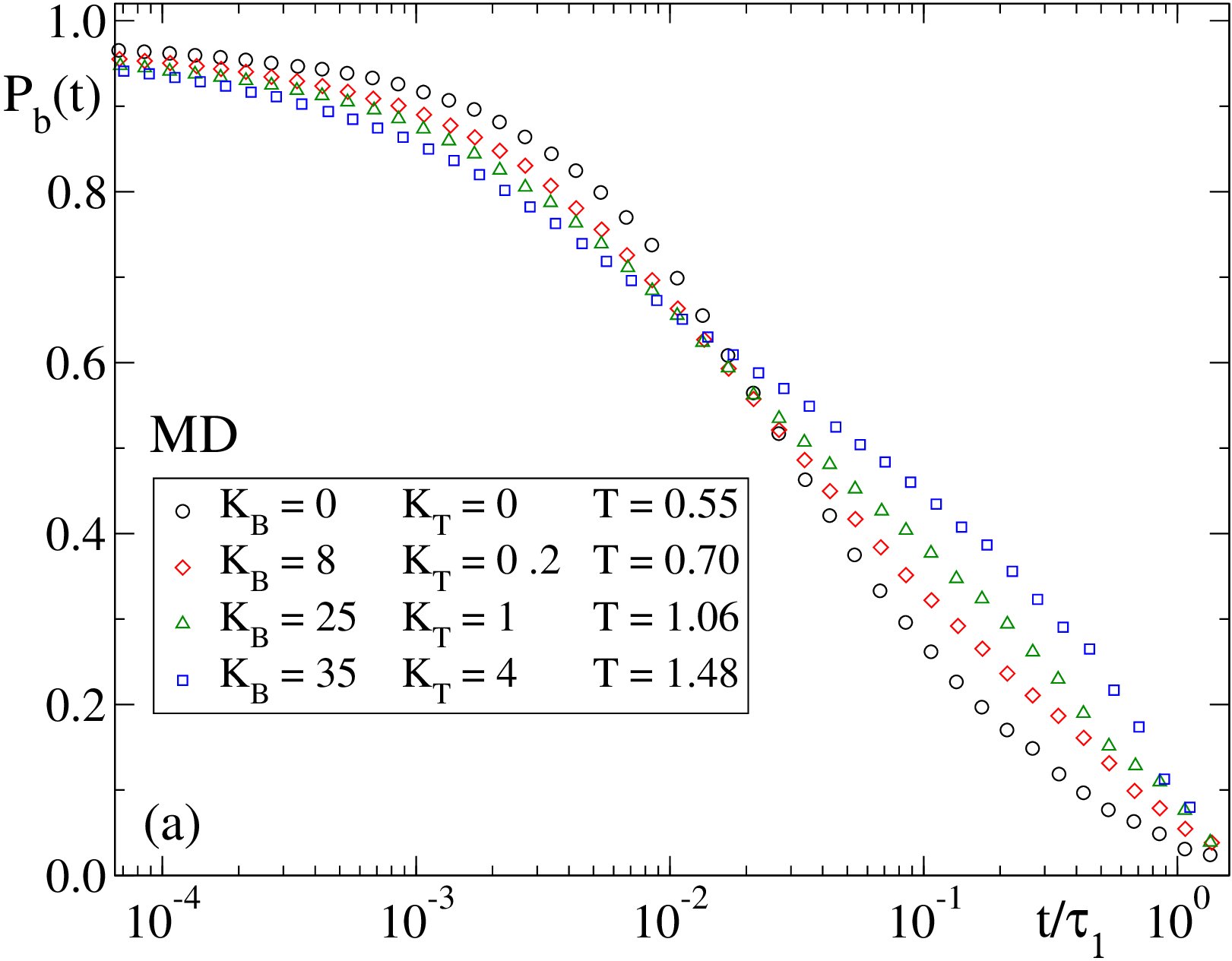}
\newline
\newline
\includegraphics[width=0.84\linewidth]{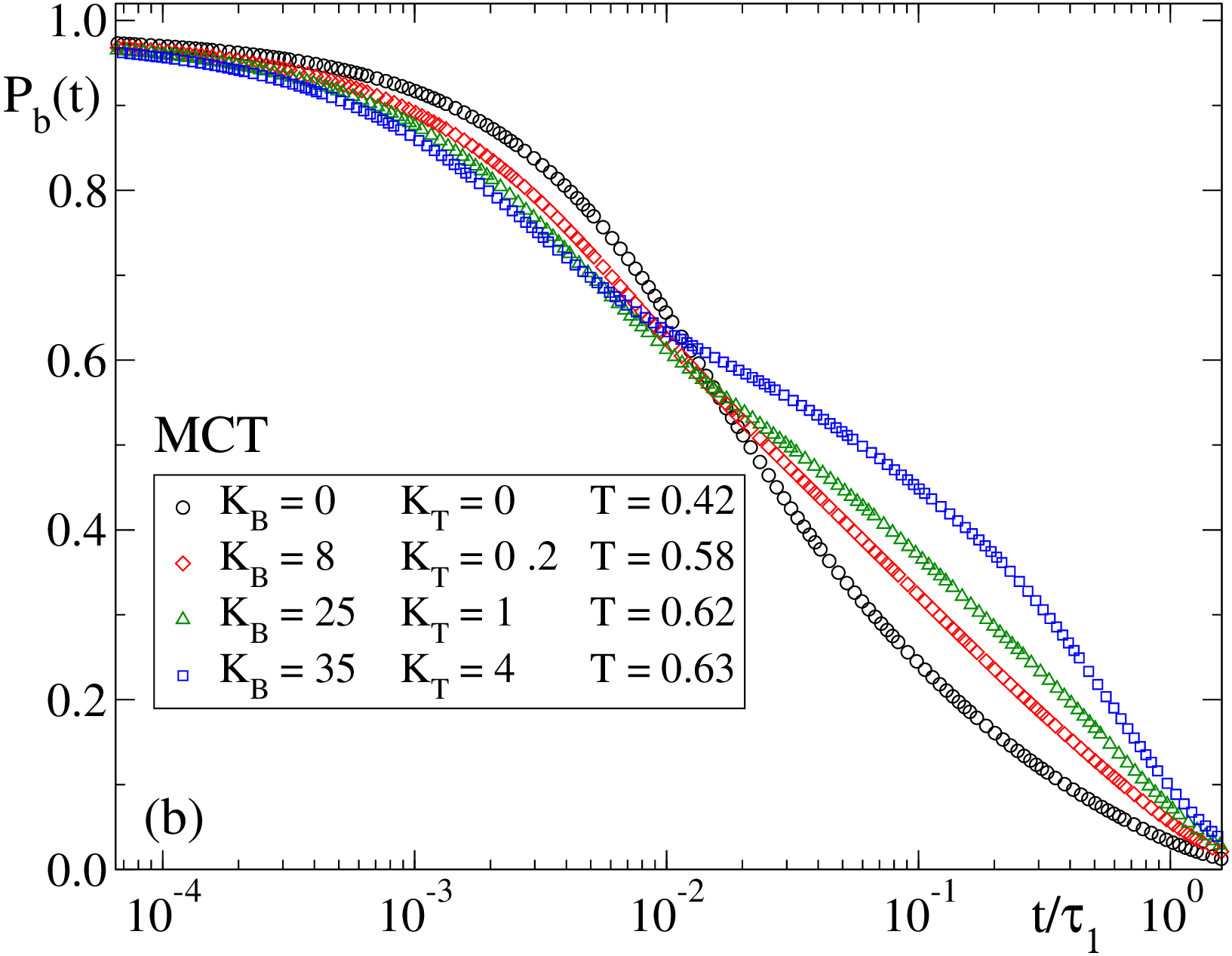}
\newline
\caption{Normalized bond correlators $P_{\rm b}(t)$ for several values
of the barrier strength at selected temperatures (see legend).
Panel (a): simulation results. Panel (b): MCT numerical solutions.
In both panels  the absolute time is rescaled by the relaxation time $\tau_1$ of the $p=1$  mode}
\label{fig:bond}
\end{center}
\end{figure}

As we observed in Fig.~\ref{fig:rousestatic} for $\Psi_{pq}(0)$, 
there are off-diagonal terms of the intrachain static 
correlations which are non-orthogonal. This non-orthogonality persists
over long time scales, as can be seen in Fig.~\ref{fig:rousecross}. The latter
shows  simulation and theoretical results for normalized Rouse cross-correlators $\Phi_{pq}(t)$, 
with $p=3$ and $q=1,3,5,7,9$. Data correspond to the same temperatures
and barrier strength (the stiffest investigated case) of the diagonal correlators of Fig.~\ref{fig:rousemodes}.
Again, MCT qualitatively reproduces simulations trends for the case of the off-diagonal terms.

Finally, it is worth noting that the good agreement between simulations and MCT 
for the Rouse correlators is similar for other observables  probing chain  dynamics.
The reason is that, through the transformation ${\bf X}_p(t) =  \sum_{j=1}^{N_{\rm m}}P_{jp}{\bf r}_j(t)$ (see above),
such observables can be  expressed in terms of the Rouse diagonal 
and cross-correlators \cite{DoiEdwards:Book1986}.
An example is given by the orientational bond correlator 
$P_{\rm b}(t)=\langle {\bf b}(0) \cdot {\bf b}(t) \rangle / \langle b^2 (0)\rangle$, 
where ${\bf b}(t)$ is the bond vector joining two consecutive monomers.
Following the former transformation we find 
$\langle {\bf b}(0) \cdot {\bf b}(t) \rangle \equiv \sum_{j=1}^{N_{\rm m}-1} \sum_{p,q=0}^{N_{\rm m}-1} 
[P^{-1}_{p,j+1}-P^{-1}_{p,j}][P^{-1}_{q,j+1}-P^{-1}_{q,j}]\langle {\bf X}_p(t) \cdot {\bf X}_q(0) \rangle$,
where $P^{-1}$ is the inverse of the matrix of coefficients $P_{jp}$.
Note that this expression is {\it exact} 
(the Rouse model makes the approximation $\langle {\bf X}_p(t) \cdot {\bf X}_q(0) \rangle =0$ for $p \ne q$).
Since MCT solutions provide the Rouse correlators for all $(p,q)$, insertion of these in the
former exact expression directly provides $P_{\rm b} (t)$.
Fig.~\ref{fig:bond} shows simulation
and MCT results of $P_{\rm b}(t)$ for several values of $(K_{\rm B}, K_{\rm T})$ from
the fully-flexible limit to the stiffest investigated chains.
As in previous figures, times are rescaled by the respective $\tau_1$,
and data in both panels correspond to a common separation parameter $\epsilon_T \approx 0.2$.
MCT reproduces semiquantitatively the observed simulation trends. These include, 
on increasing  barrier strength,
a relative speed up and slowing down (in terms of the {\it scaled} time $t/\tau_1$) 
of respectively the short-time and long-time dynamics. MCT also accounts for
the emergence, for strong barriers, of a plateau at $t/\tau_1 \sim 10^{-2}$
and a change in the concavity of the decay.

\begin{center}
{\bf IV. DISCUSSION}
\end{center}

On Section III we have shown that, concerning the critical temperature $T_{\rm c}$, 
MCT reproduces qualitative simulation trends for low and moderate barriers. 
However a strong disagreement is found on  approaching the limit
of stiff chains. We have also found a clear discrepancy in the trends of the 
$\lambda$-exponent, with a nearly constant value from theory
and strongly barrier-dependent values from simulations. 
In this Section we discuss possible origins of these discrepancies.

\begin{center}
{\bf A. Three-point static correlations}
\end{center}

In Ref.~\cite{Bernabei:JCP2009} we showed that the failure of the MCT predictions
for strong intramolecular barriers was not apparently related with the breakdown
of the PRISM approximations, which are invoked in the derivation of the MCT equations for polymers.
Indeed the quality of such approximations appeared to be the same for all the
range of barrier strength here investigated by simulation and MCT.
Despite the mentioned discrepancies between theory and simulation,
the phenomenological analysis of simulation results in terms of a huge set of general asymptotic
laws of MCT was consistent \cite{Bernabei:PRL2008,Bernabei:JCP2009}. 
This means that the dynamic exponents involved in the different
tested laws could be, in each case, related to a same $\lambda^{\rm MD}$. As we discussed
in \cite{Bernabei:JCP2009}, such scaling laws are a mathematical consequence of the bilinear dependence
of the memory kernel on the density correlators (see Eq.~(\ref{eq:mMCT})). The specific numerical
values of $\lambda$ (and by transformation, of the other dynamic exponents) are determined by the
static quantities entering the vertex (\ref{eq:vertMCT}) \cite{Franosch:PRE1997,Bernabei:JCP2009}. 
Given the consistency of the
phenomenological analysis we speculated that, by retaining the bilinear form of the MCT memory kernel,  
there may be missing static contributions in the vertex which are not significant
for low barriers, but become increasingly important as the limit of stiff chains is approached.
Including them and solving the MCT equations accordingly, might raise
the theoretical values of  $T_{\rm c}$ and $\lambda$, leading to a better agreement
with the simulation trends.

Thus, we suggested  that {\it intrachain} three-point static correlations
should be explicitly included in the MCT vertex. Chain stiffness induces a strong directionality
in the intrachain static correlations, at least at near-neighbor distances. 
It has been shown that directionality in static correlations
can break the static convolution approximation of MCT, Eq.~(\ref{eq:convol}). 
A well-known example is given by silica, a network-forming system. 
For the latter the inclusion of three-point static correlations in the MCT vertex significantly improves
the comparison between theory and simulation, with respect to the solutions obtained
under the convolution approximation \cite{Sciortino:PRL2001}.

The calculation of the three-point static correlations involved
in Eq.~(\ref{eq:convol}) is very demanding.  This is because most of the computational time
is consumed by the {\it interchain} three-point correlations.
For intrachain three-point correlations the computation is not demanding.
Fortunately, in the present case only the latter is necessary, since 
the directionality of correlations is only relevant along the chain.
Thus, the convolution approximation is retained for interchain correlations,
and it is modified only to include the intrachain three-point correlations.
With this, the new MCT vertex reads \cite{chongw3}
\begin{eqnarray}
{\cal V}({\bf q},{\bf q}-{\bf k}) = \frac{\rho}{2q^4}S(q)S(k)S(|{\bf q}-{\bf k}|)[{\bf q}\cdot{\bf k}c(k) \nonumber \\
  + {\bf q}\cdot ({\bf q}-{\bf k})c(|{\bf q}-{\bf k}|)+
\rho q^2c_3({\bf q},{\bf q}-{\bf k})]^2 ,
\label{eq:vertMCTc3}
\end{eqnarray}
where $c_3({\bf q},{\bf q}-{\bf k})$ is the three-point {\it intramolecular}  direct correlation function, given by
\begin{eqnarray}
\rho^2 c_3({\bf q},{\bf q}-{\bf k})=1-\frac{\omega_3({\bf q},{\bf q}-{\bf k})}{\omega(q)\omega(k)\omega(|{\bf q}-{\bf k}|)},
\label{eq:c3}
\end{eqnarray}
and $\omega_3({\bf q},{\bf q}-{\bf k})$ is the three-point intramolecular structure factor
\begin{eqnarray}
\omega_3({\bf q},{\bf q}-{\bf k}) = 
\frac{1}{N_{\rm c}N_{\rm m}} \sum_{I=1}^{N_{\rm c}}\sum_{a,b,c=1}^{N_{\rm m}} \times \nonumber \\
\exp\{i[-{\bf q}\cdot{\bf r}_a^I+{\bf k}\cdot{\bf r}_b^I +({\bf q}-{\bf k})\cdot{\bf r}_c^I]\} .
\label{eq:w3}
\end{eqnarray}
Indices in (\ref{eq:w3}) are defined as in (\ref{eq:wq}).
The convolution approximation for intrachain correlations assumes 
$\omega_3({\bf q},{\bf q}-{\bf k}) = \omega(q)\omega(k)\omega(|{\bf q}-{\bf k}|)$,
or equivalently $c_3({\bf q},{\bf q}-{\bf k}) =0$, reducing the vertex (\ref{eq:vertMCTc3}) 
to the original Eq.~(\ref{eq:vertMCT}).

\begin{figure}
\begin{center}
\includegraphics[width=0.82\linewidth]{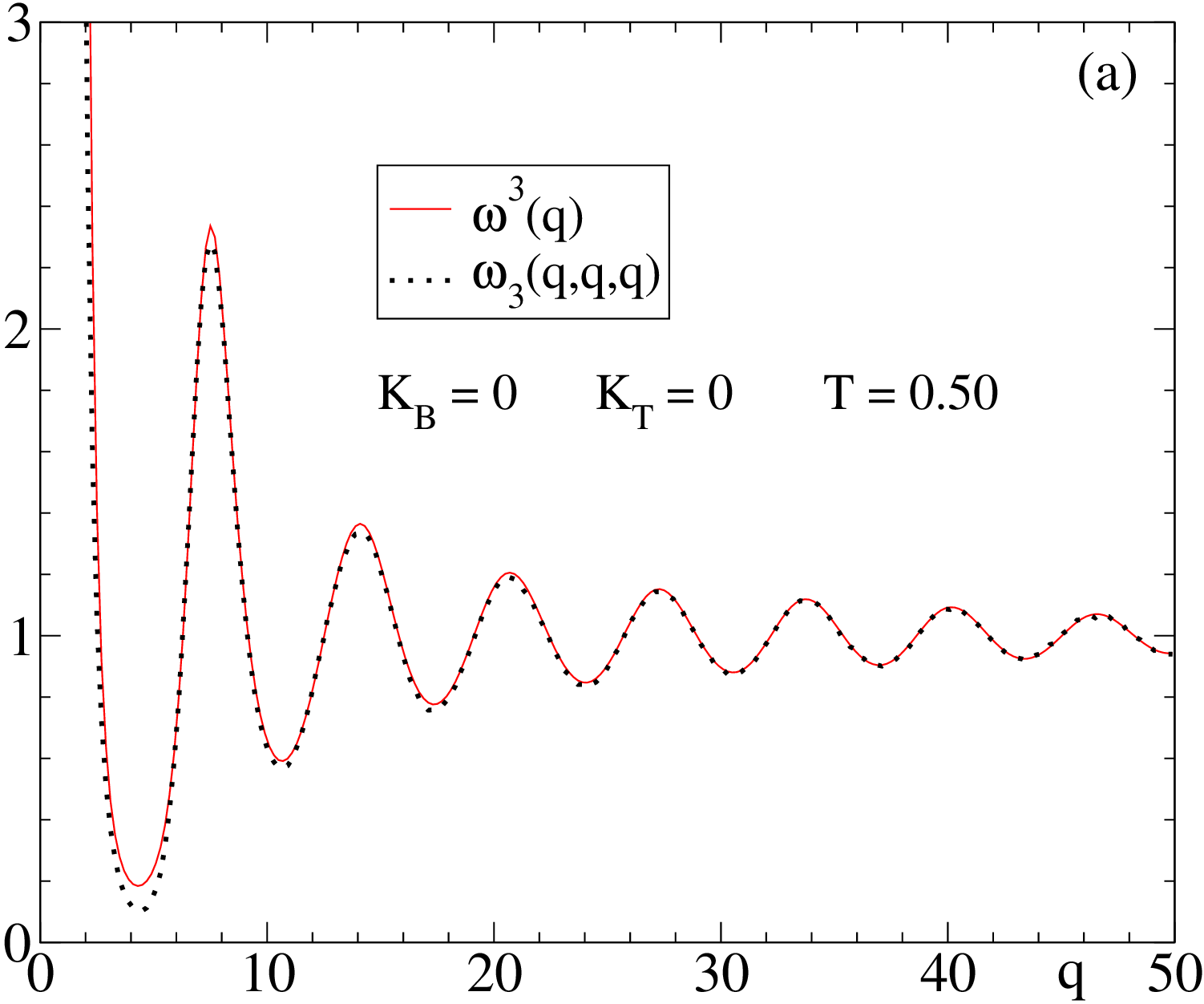}
\newline
\newline
\includegraphics[width=0.82\linewidth]{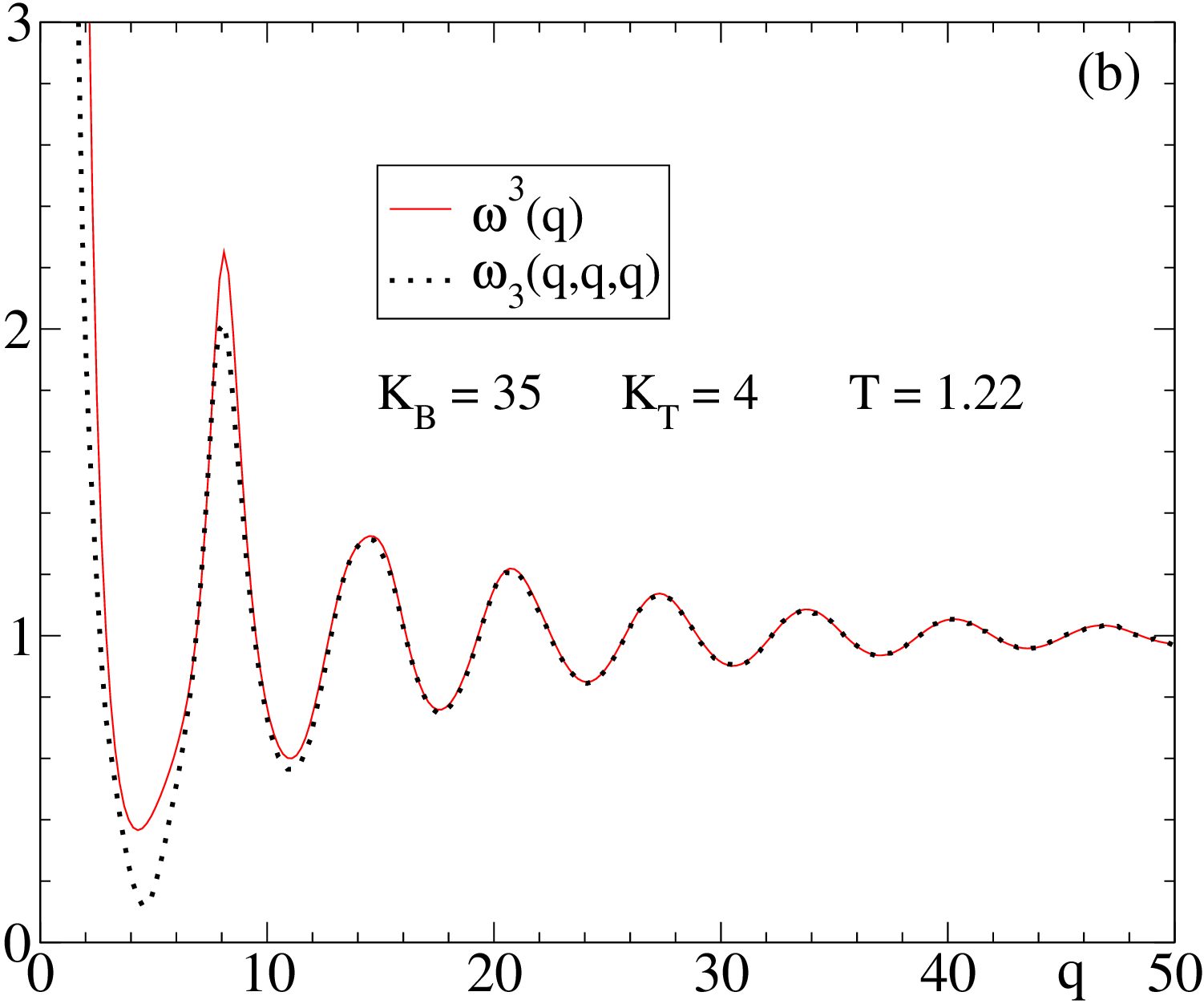}
\newline
\caption{Comparison between $\omega_3(q,q,q)$ (dotted lines) and the convolution
approximation $\omega^3(q)$ (solid lines). Panel (a): fully-flexible chains at $T = 0.50$.
Panel (b): Stiffest investigated chains at $T = 1.22$.}
\label{fig:w3flex}
\end{center}
\end{figure}

\begin{figure}
\begin{center}
\includegraphics[width=0.82\linewidth]{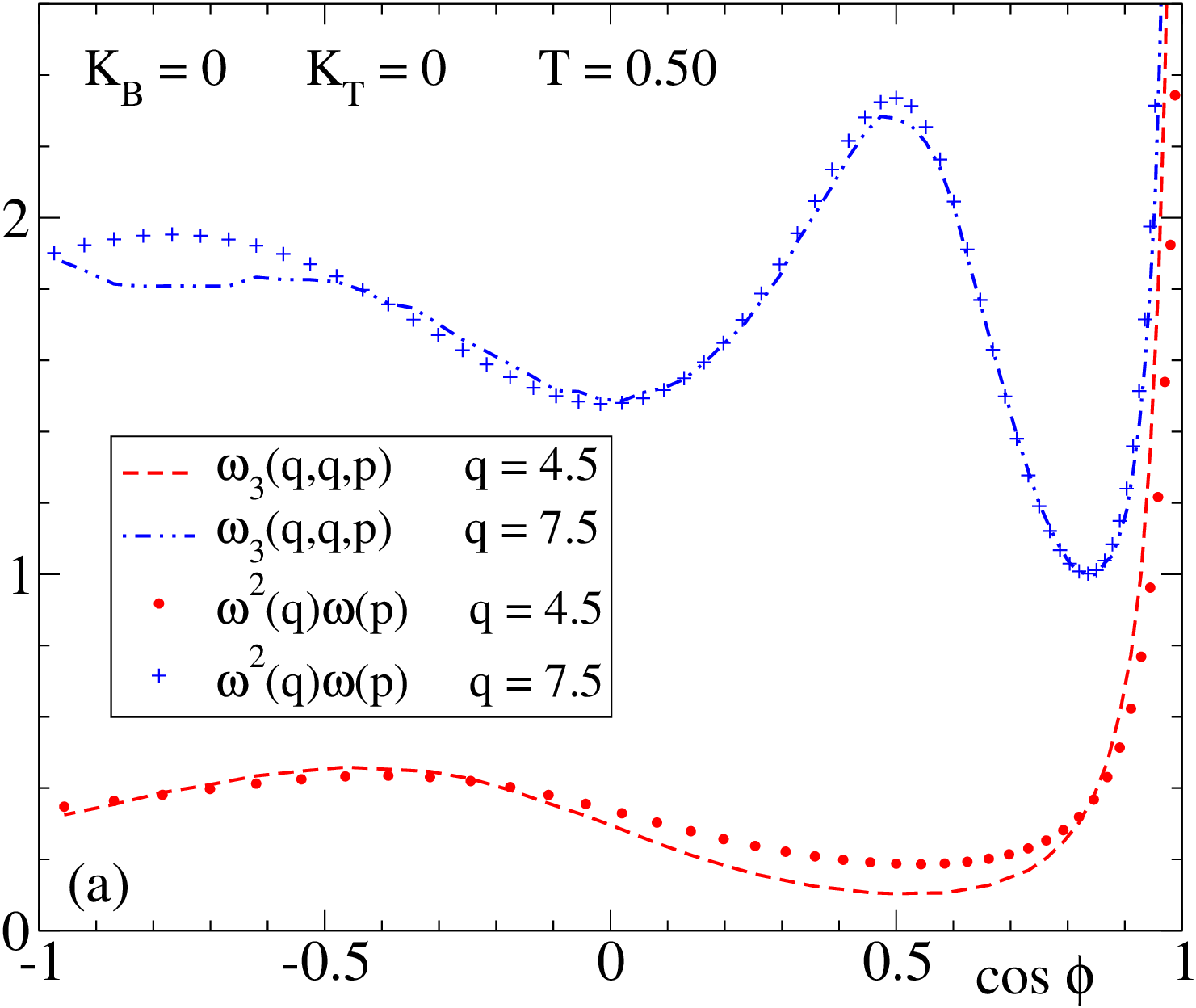}
\newline
\newline
\includegraphics[width=0.82\linewidth]{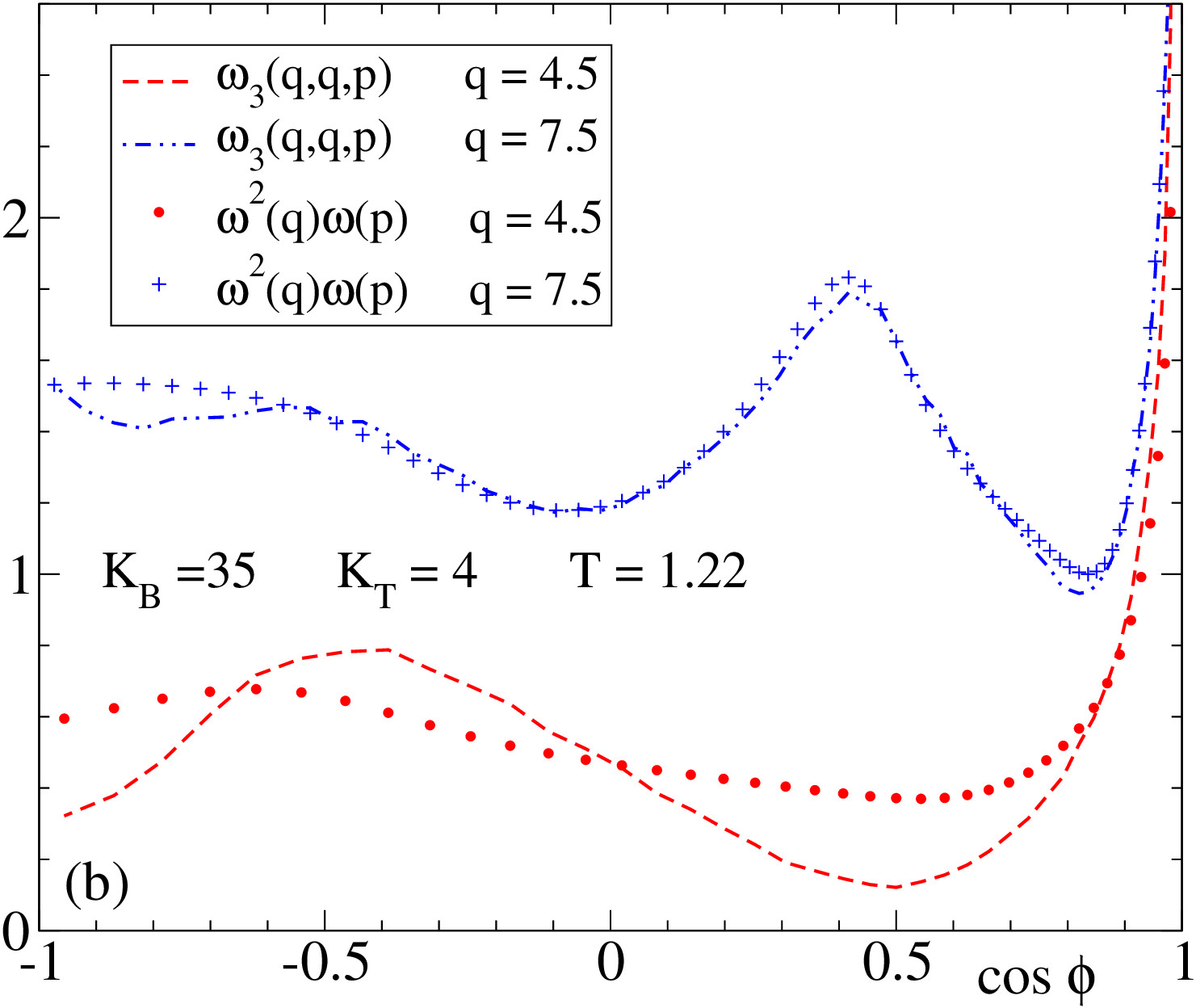}
\newline
\caption{Comparison of $\omega_3(q,q,p=q\sqrt{2(1-\cos{\phi})})$ (lines)
with the convolution approximation $\omega^2(q)\omega(p)$ (symbols), for two
selected values $q = 4.5$ and 7.5.
Panel (a): fully-flexible chains at $T = 0.50$.
Panel (b): Stiffest investigated chains at $T = 1.22$.}
\label{fig:w3stiff}
\end{center}
\end{figure}

Figs.~\ref{fig:w3flex} and \ref{fig:w3stiff} show representative tests of the convolution
approximation for respectively fully-flexible and stiffest investigated chains.
Following the scheme proposed in Ref.~\cite{Chong:PRE2007}, 
the vectors ${\bf q}$, ${\bf k}$ and ${\bf p}={\bf q}-{\bf k}$ define the sides of a triangle,
the first two enclosing an angle $\phi$ given by $\cos{\phi}=(q^2+k^2-p^2)/2qk$.
Panels (a) and (b) in Fig.~\ref{fig:w3flex}  show a test of the corresponding expression for an equilateral triangle,
$\omega_3(q,q,q) = \omega^3(q)$.
Panels (a) and (b) in Fig.~\ref{fig:w3stiff}  show a similar test for equal moduli $k = q$
and all the relative orientations (given by $\cos\phi$) of $\bf{q}$ and $\bf{k}$.
In other words, we test  the approximation $\omega_3(q,q,p=q\sqrt{2(1-\cos{\phi})})= \omega^2(q)\omega(p)$.
Data in Fig.~\ref{fig:w3stiff}  are represented as a function of $\cos\phi$ for two characteristic
wave vectors, corresponding to the first minimum and second maximum of the respective $\omega_3(q,q,q)$
(see Fig.~\ref{fig:w3flex}).

As already noted in Ref.~\cite{Aichele:PRE2004}, the convolution approximation  for intrachain static correlations
provides a good description of $\omega_3$ in the fully-flexible limit. As expected, the quality
of the approximation decreases by introducing intramolecular barriers. 
Still it constitutes a good approximation for all the investigated barrier strength. 
In the case of wave vectors around the first peak of  $S(q)$, $q_{\rm max} \approx 7$, 
the quality is almost unaffected by the barrier strength, i.e., 
the terms $c_3({\bf q}_{\rm max},{\bf q}_{\rm max}-{\bf k})$ will be small even for the stiffest investigated chains.
It must be noted that the MCT kernel is usually dominated by the contributions around $q_{\rm max}$.
Thus, the former observations suggest that the inclusion of the three-point static correlations
will modify weakly the MCT solutions obtained under the convolution approximation. 
We confirm this by obtaining numerical solutions with the vertex (\ref{eq:vertMCTc3}),
for which we compute the input quantities involved in Eqs.~(\ref{eq:c3}, \ref{eq:w3}) directly from the simulations.
The so-obtained values of the critical temperature $T_{\rm c}$ and $\lambda$-exponents raise by
$\approx 1\%$ as much, even for the stiffest chains, with respect to the
previous values (Fig.~\ref{fig:lambdaTc}) found under the assumption $c_3 =0$.
With all this, we conclude that the observed discrepancies between simulation
and theoretical trends of $T_{\rm c}$ and $\lambda$ are not related to the breakdown
of the convolution approximation for static three-point correlations. The latter indeed retains
its validity for all the investigated range of barrier strength.

\begin{center}
{\bf B. Dynamic heterogeneities}
\end{center}

It is well-known that the quality of the Kawasaki approximation for dynamic correlations
(see above) breaks on decreasing temperature. This feature is specially critical around the $\alpha$-time scale
\cite{Glotzer:JNCS2000,Lacevic:JCP2003,Berthier:Science2005,Szamel:PRE2006,Berthier:JCP2007},
leading to the complete failure of the MCT predictions associated to it, as the power law behavior
$D^{-1}, \tau_{\alpha} \sim (T-T_{\rm c})^{-\gamma}$, or the time-temperature superposition of density correlators.
This breakdown is usually assigned to the emergence of strong dynamic heterogeneities
on approaching the glass transition 
\cite{Glotzer:JNCS2000,Lacevic:JCP2003,Berthier:Science2005,Szamel:PRE2006,Berthier:JCP2007}.
Having noted this we may speculate that, for some reason to be understood, 
increasing the barrier strength strongly enhances dynamic heterogeneities.
This might result in a lower quality of the MCT and might be the reason for the observed discrepancies between
theory and simulation trends for $T_{\rm c}$. 

Now we show that this is not actually the case, and that there is no correlation
between barrier strength and enhanced dynamic heterogeneity. 
Non-gaussian parameters provide a simple way of quantifying the strength of the dynamic heterogeneity.
They display large positive values at the time scales for which the respective van Hove function 
strongly deviates from the gaussian limit. This occurs when
a significant fraction of particles has performed displacements very different from the average.
Here we discuss the two most popular non-gaussian parameters.
The standard or `fast' parameter is given by
$\alpha_2(t)=\frac{3}{5}\langle (\Delta r(t))^4 \rangle /\langle (\Delta r(t))^2 \rangle^2 -1$.
The `slow' parameter introduced by Flenner and Szamel \cite{Flenner:PRE2005}
is defined as $\gamma_2(t)=\frac{1}{3}\langle (\Delta r(t))^2 \rangle \langle 1/(\Delta r(t))^2 \rangle -1$.

By construction $\alpha_2(t)$ and $\gamma_2 (t)$ are identically zero 
for a gaussian form of the van Hove self-correlation function. However, as noted in \cite{Flenner:PRE2005},
large positive values of these parameters have a very different microscopic origin, reflecting
distinct aspects of dynamic heterogeneity. In the case of the fast parameter $\alpha_2 (t)$,
large values originate from a significant population of particles which have performed much larger
displacements than the average. This effect is generally maximum at the time scale $t^{\ast}$ 
around the end of the caging regime. Thus, $\alpha_2(t)$ increases from zero at $t =0$
up to a maximum at $t^{\ast}$, and decays to zero at longer times. The increase of 
the maximum $\alpha_2(t^{\ast})$ on decreasing temperature reflects a progressive
enhancement of dynamic heterogeneity, at the decaging process, on approaching the glass transition.

The slow parameter $\gamma_2(t)$ exhibits analogous trends for the temperature and time-dependence.
However the maximum of $\gamma_2(t)$ takes place at much longer scales than $t^{\ast}$,
namely around the $\alpha$-relaxation time $\tau_{\alpha}$. This effect
originates from a significant population of particles which at $t \sim \tau_{\alpha}$ have
performed much {\it smaller} displacements than the average \cite{Flenner:PRE2005}.

\begin{figure}
\includegraphics[width=0.86\linewidth]{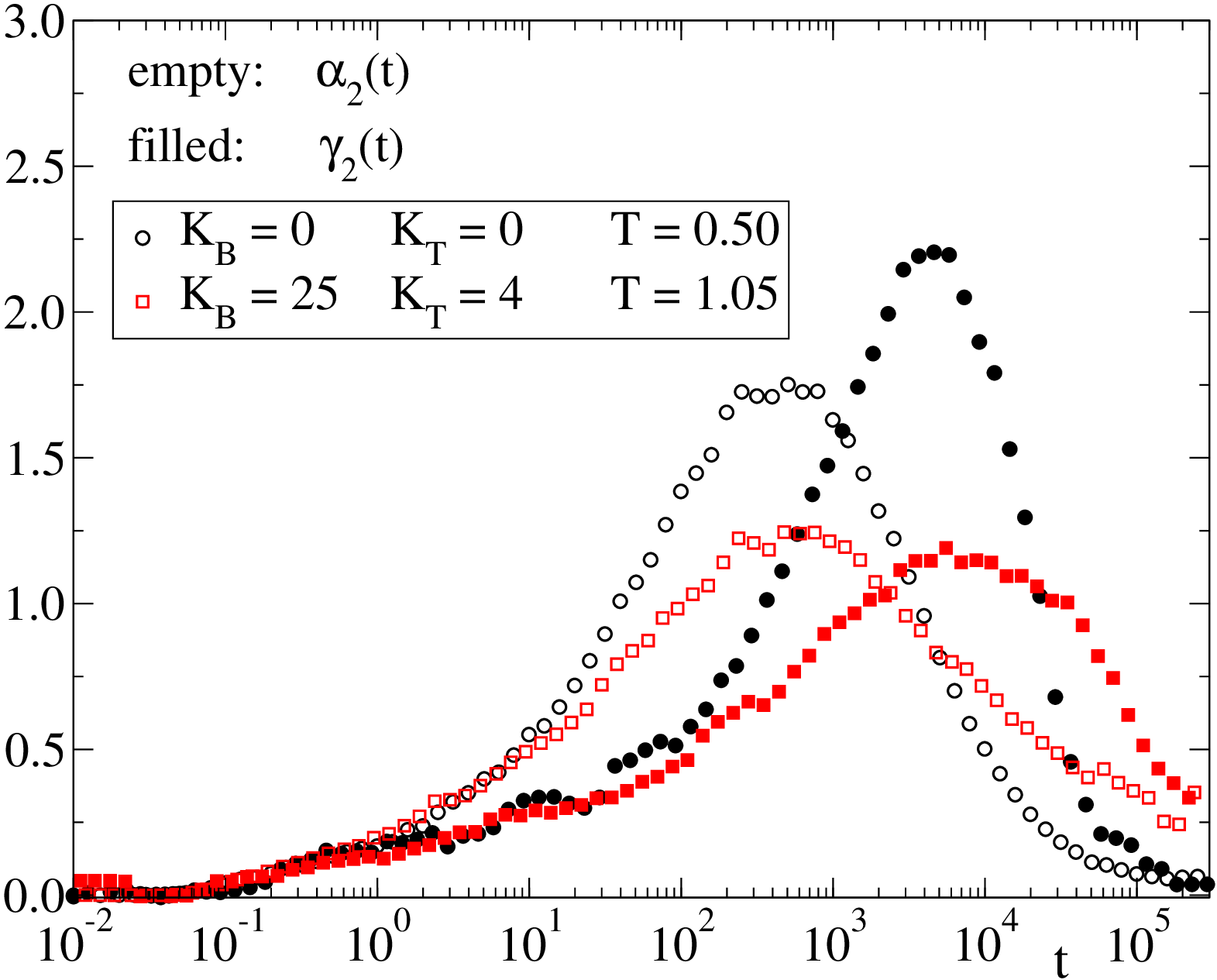}
\caption{For two selected temperatures with similar decaging and $\alpha$-relaxation times,
non-gaussian parameters of fully-flexible (circles) and stiff chains with 
$(K_B,K_T)=(25,1)$ (squares). Empty and filled symbols correspond respectively
to the fast ($\alpha_2(t)$) and slow ($\gamma_2(t)$) parameters.} 
\label{fig:nongauspar}
\end{figure}

\begin{figure}
\includegraphics[width=0.86\linewidth]{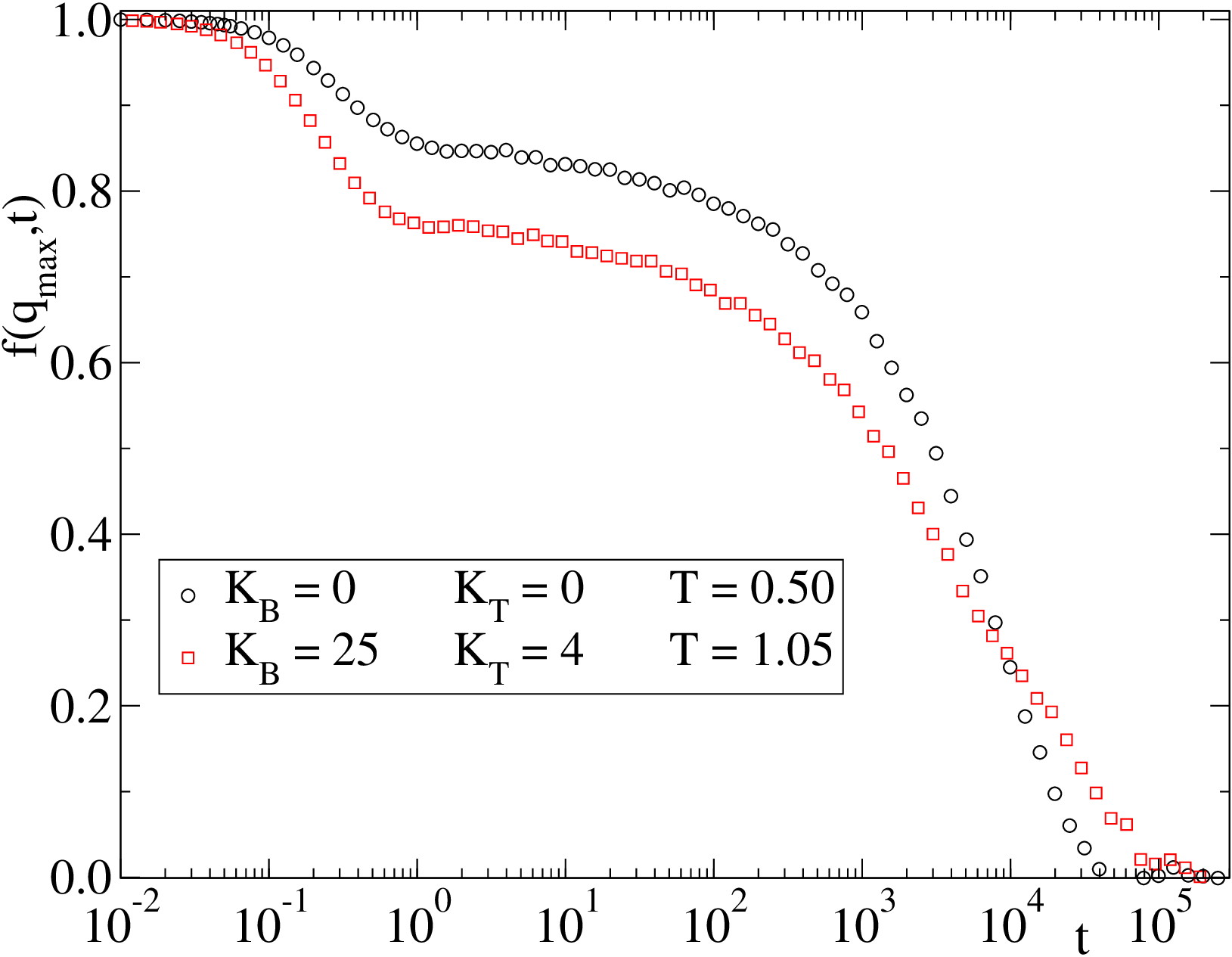} 
\caption{Density-density correlator $f(q_{\rm max},t)$
for fully-flexible (circles) and stiff chains with 
$(K_B,K_T)=(25,1)$ (squares). 
Data are shown for the same selected temperatures of Fig.~\ref{fig:nongauspar}.} 
\label{fig:compfcoh2}
\end{figure}

Fig.~\ref{fig:nongauspar} compares simulation
results of $\alpha_2(t)$ and $\gamma_2(t)$,  for the fully-flexible case
and for very stiff chains. For a fair comparison we have selected
temperatures at which the respective $\alpha$-relaxation times are similar.
These are $T = 0.50$ and $T = 1.05$, for respectively fully-flexible and stiff chains,
and correspond to a separation parameter $\epsilon_T \sim 0.04$ (see Fig.~\ref{fig:lambdaTc}).
Let us remind that the $\alpha$-time scale can be estimated, e.g., as $f(q_{\rm max},t) =0.2$.
In Fig.~\ref{fig:compfcoh2} we display $f(q_{\rm max},t)$ for both systems
at the former temperatures, showing that the respective $\alpha$-time scales are roughly the same,
$\tau_{\alpha} \sim 10^4$. The decaging times, which can be estimated
from the start of the decay from the plateau in  $f(q_{\rm max},t)$, are also roughly the same,
$t^{\ast} \sim 500$.
This equivalence is indeed reflected in the trends of the non-gaussian parameters in Fig.~\ref{fig:nongauspar}.
Thus, in both systems $\alpha_2 (t)$ is  peaked at $t^{\ast} \sim 500$ and  $\gamma_2 (t)$
is peaked at $\tau_{\alpha} \sim 10^4$.

Having noted this equivalence of time scales, data in Fig.~\ref{fig:nongauspar} do not reflect any enhancement
of dynamic heterogeneity on increasing the barrier strength. Actually, the opposite
effect is suggested by the lower values of $\alpha_2(t)$ and $\gamma_2 (t)$ 
for stiff chains with respect to the fully-flexible case. With this, we discard a major role of dynamic
heterogeneities as the reason for the observed discrepancies between simulations and MCT solutions
for very stiff chains.

\begin{center}
{\bf C. Chain packing}
\end{center}

As mentioned in Section III and shown in Fig.~\ref{fig:tauq},
MCT fails at reproducing the peak around $q_{\rm C} \sim 4$
for the $q$-dependence of the KWW times of density-density correlators.
As noted in Ref.~\cite{Chong:PRE2007} for the fully-flexible case, 
the origin of this peak may be related
to dynamic correlations between centers-of-mass of the chains. The latter might arise
from the effective packing between the polymer coils, interacting as fully penetrable
spheres of size $R_{\rm g}$. This interpretation is not clear in view of the results
of Fig.~\ref{fig:tauq}, since the value of $q_{\rm C}$ does not seem to be related with $2\pi/R_{\rm g}$.
Having noted this, Chong {\it et al.} found that the incorporation of the static
correlations between the centers-of-mass in the MCT equations did not improve
the description of the simulation results. As shown in \cite{Chong:PRE2007}, this is not unexpected
due to the almost featureless form of the static structure factor of the centers-of-mass
$S_{\rm CM}(q)$. The inset of Fig.~\ref{fig:tauq}a shows simulation results of $S_{\rm CM}(q)$,
for the same barrier strength and temperatures of the KWW times in the main panel.
The introduction of chain stiffness does not induce significant features in $S_{\rm CM}(q)$,
appart from a stronger signal at low $q$. The latter indeed suggests that
packing effects between the polymer coils are even weaker than for the fully-flexible case.
Within the former interpretation, this would be consistent with the lower intensity
of the mentioned peak of $\tau^{\rm K}_q /\tau_{q_{\rm max}}$ 
at $q_{\rm C}\sim 4$.
All these results suggest that discrepancies between simulations and MCT
on increasing chain stiffness are not related to a dynamic coupling, not accounted for within the theory,
to the slow modes at $q_{\rm C}\sim 4$. Indeed, this coupling seems to be weaker for stiff chains.

\begin{center}
{\bf D. Outlook}
\end{center}

In summary, in this section we have discussed possible origins for the discrepancies,
concerning the structural relaxation, between simulations and MCT on increasing barrier strength.
We discard a major role, in comparison with the fully-flexible case, of three-point static correlations,
dynamic heterogeneities and chain packing. These effects become even weaker on increasing
chain stiffness. We remind that such effects are indeed neglected 
in the derivation of the MCT equations used here (see Section III).
Results in this section suggest that this is  not less justified for very stiff chains
than for fully-flexible ones.

How to improve the theory to account for dynamic trends in stiff chains
is an open question. A way might be the reformulation or extension
of  the MCT equations, retaining the bilinear form of the kernel, in terms of new dynamic observables 
coupled to density fluctuations. 
Such observables can be adequate for describing particular dynamic features which 
are not captured by the usual observables, i.e., the number density fluctuations $\rho({\bf q},t)$.
Some examples are roto-translational site fluctuations adapted to the molecular symmetry,
as has been shown, e.g., for dumbbell-like molecules \cite{Chong:PRE2002,Chong:PRE2002b} 
or for a simple model of orthoterphenyl \cite{Chong:PRE2004}. The inclusion of density fluctuations 
of centers-of-mass  improve results for rigid molecules \cite{Chong:PRE2004}
concerning a peak in $\tau^{\rm K}_q /\tau_{q_{\rm max}}$ at intermediate $q$  \cite{Rinaldi:PRE2001}, 
similar to that observed here at $q_{\rm C} \sim 4$.
As discussed above, this is not the
case for polymer chains. Though there is no characteristic symmetry in polymer chains,
roto-translational density fluctuations can also be defined over sites $a,b$ at some 
characteristic distance $|a-b|$, perhaps probing the relevant length scale $2\pi/q_{\rm C}$,
which according to the data of Fig.~\ref{fig:tauq} seems to depend weakly on the barrier strength.
Whether this procedure may improve the agreement between MCT and simulations 
remains to be solved.

\begin{center}
{\bf V.CONCLUSIONS}
\end{center}

By means of simulations and solution of the equations of the Mode Coupling Theory, 
we have studied the role of intramolecular barriers, of arbitrary strength, 
on several aspects of polymer dynamics. 
The investigated dynamic range extends from the caging regime characteristic of glass-formers
to the relaxation of the chain Rouse modes.
Solutions of the MCT for the structural relaxation reproduce qualitative trends of simulations for weak
and moderate barriers. However, a progressive discrepancy between MCT and simulations is revealed as
the limit of stiff chains is approached. We have tested the validity of several assumptions inherent to
the theory. Deviations from the theoretical predictions do not seem related with  
dynamic heterogenities, which indeed are not enhanced by increasing the barrier strength.
Moreover, the convolution approximation for three-point static correlations 
retains its validity for stiff chains. Even the role of slow modes at intermediate length scales, not
accounted for by MCT, becomes less significant on increasing chain stiffness.
At this point it is not clear how to improve the MCT equations in order to remove the mentioned discrepancies
for the case of stiff chains.
We have suggested the possibility of formulating the MCT equations
in terms of roto-translational density fluctuations over specific length scales.

Concerning the relaxation of the chain degrees of freedom, MCT provides a microscopic basis for 
the observed deviations from the Rouse model on increasing the barrier strength.
These include anomalous scaling of relaxation times,
long-time plateaux, and non-monotonous wavelength dependence of the mode correlators.
Beyond usual phenomenological models for chain dynamics (the Rouse model being the corresponding
one for fully-flexible chains), MCT provides a unified microscopic picture  
down to time scales around and before the $\alpha$-process, which 
is not accounted for within the mentioned models.

\begin{center}
{\bf ACKNOWLEDGEMENTS}
\end{center}

We thank S. -H. Chong, T. Franosch, M. Fuchs, M. Sperl, and J. Baschnagel for useful discussions.
We acknowledge financial support from the projects
FP7-PEOPLE-2007-1-1-ITN (DYNACOP, EU),
MAT2007-63681 (Spain), IT-436-07 (GV, Spain), ERC-226207-PATCHYCOLLOIDS (EU), 
and ITN-234810-COMPLOIDS (EU).
%


\begin{thebibliography}{58}
\bibitem[{not({\natexlab{b}})}]{notebreak}
\bibinfo{note}{The breakdown of ideal MCT on approaching the glass transition
  temperature $T_{\rm g}$ is reflected in the failure of the asymptotic laws.
  In particular, for very small values of the separation parameter, typically
  $\epsilon_T \sim 10^{-2}$, a crossover takes place from power-law to
  Vogel-Fulcher-Tamman behavior, $\tau_{\alpha} \sim \exp[A/(T-T_0)]$ (or
  Arrhenius-like, $T_0 = 0$, in strong liquids), which extends down to $T_{\rm
  g}$. Still, there is usually a range of about two decades in $\epsilon_T$ for
  which the predicted asymptotic laws are observed.}
\bibitem[{not({\natexlab{d}})}]{notex}
\bibinfo{note}{E.g., for $(K_{\rm B}, K_{\rm T})=(25,1)$ we find a variation of
  $\approx 7 \%$ in $x$ for a change in $T$ by a factor $\sim 2$.}
\end{thebibliography}

\end{document}